\documentclass[journal]{IEEEtran}

 \usepackage{amsmath, amssymb, bm, subfigure ,  graphicx}
 \usepackage{algorithm}
 \usepackage{hyperref}
\usepackage[noend]{algorithmic}

\usepackage{graphicx}

\usepackage{graphicx}
\usepackage{caption}
\usepackage[usenames,dvipsnames]{xcolor}
\usepackage{listings}

\lstdefinestyle{customc}{
  belowcaptionskip=1\baselineskip,
  breaklines=true,
  xleftmargin=\parindent,
  language=MATLAB,
  showstringspaces=false,
basicstyle=\footnotesize\ttfamily,
  keywordstyle=\bfseries\color{green!40!black},
  commentstyle=\itshape\color{purple!40!black},
  identifierstyle=\color{blue},
  stringstyle=\color{orange},
  }

\lstdefinestyle{customasm}{
  belowcaptionskip=1\baselineskip,
  frame=L,
  xleftmargin=\parindent,
  language=[x86masm]Assembler,
  basicstyle=\footnotesize\ttfamily,
  commentstyle=\itshape\color{purple!40!black},
}

\newtheorem{theorem}{Theorem}
\newtheorem{lemma}{Lemma}

\newtheorem{remark}{Remark}

\newcommand{\by}{\mathbf{y}}
\newcommand{\bx}{\mathbf{x}}
\newcommand{\bz}{\mathbf{z}}

\newcommand{\ba}{\mathbf{a}}

\newcommand{\omegahat}{\hat{\omega}}
\newcommand{\delomega}{\Delta\omega_{\text{min}}}
\newcommand{\ghat}{\hat{g}}

\newcommand{\parSet}{\mathcal{P}}
\newcommand{\byres}{\by_{\text{r}}}
\newcommand{\norm}[1]{\left\Vert{#1}\right\Vert}
\newcommand{\CRB}{Cram\'{e}r Rao Bound}
\newcommand{\DDFT}{\Delta_{\text{dft}}}

\newcommand{\ignore}[1]{}

\newcommand*\Function[2]{\STATE\textbf{Procedure} \textsc{#1}(#2):}

\newcommand*\While[1]{\WHILE{#1}}
\newcommand*\EndWhile{\ENDWHILE}
\newcommand*\Return[1]{\textbf{return} #1}
\newcommand*\State[1]{\STATE #1}
\newcommand*\EndFunction{}

\newcommand{\minSep}{\Delta\omega_{\min}}

\author{\IEEEauthorblockN{B. Mamandipoor, D. Ramasamy, U. Madhow }\\
\IEEEauthorblockA{ECE Department, University of California Santa Barbara\\
\{bmamandi, dineshr, madhow\}@ece.ucsb.edu}}

\title{Newtonized Orthogonal Matching Pursuit: Frequency Estimation over the Continuum}
\date{}

\begin{document}

\maketitle

\begin{abstract}
We propose a fast sequential algorithm for the fundamental problem of estimating frequencies and amplitudes of a noisy mixture of sinusoids. The algorithm is a natural generalization of Orthogonal Matching Pursuit (OMP) to the continuum using Newton refinements, and hence
 is termed Newtonized OMP (NOMP). Each iteration consists of two phases: detection of a new sinusoid, and sequential Newton refinements of the parameters of already detected sinusoids. The refinements play a critical role in two ways: (1) sidestepping the potential \emph{basis mismatch} from discretizing a continuous parameter space, (2) providing \emph{feedback} for locally refining parameters estimated in previous iterations. 
We characterize convergence, and provide a Constant False Alarm Rate (CFAR) based termination criterion.  
 By benchmarking against the \CRB, we show that NOMP achieves near-optimal performance under a variety of conditions. 
We  compare the performance of NOMP with classical algorithms such as MUSIC 
and more recent Atomic norm Soft Thresholding (AST) and Lasso algorithms, both in terms of frequency estimation accuracy and run time.
\end{abstract}



\section{Introduction}
Frequency estimation from a mixture of sinusoids in AWGN is a fundamental problem that arises in a variety of communication and radar applications, including estimation of spatial channels (e.g., for phased arrays), temporal multipath channels (e.g. for equalization), and spatiotemporal channels (e.g., range and direction of arrival estimation for a target). In all of these applications, the spectrum of the measured signal consists of multiple discrete frequencies over a continuous interval. The problem of estimating the frequencies of such a signal is also known as line spectral estimation.

In this paper, we propose an algorithm to estimate frequencies from $N$ equi-spaced noisy samples in time, denoted by $\mathbf{y} \in \mathbb{C}^{N}$. 
Defining the unit norm sinusoid of frequency $\omega \in [0,2\pi)$, by
\begin{equation}
\bx(\omega) \triangleq {1\over\sqrt{N}}~ [1~e^{j\omega}~\cdots~e^{j(N-1)\omega}]^T,
\end{equation}
the observed signal is a mixture of $K$ sinusoids: 
\begin{equation}
\by =\sum_{l = 1}^{K} g_l \bx(\omega_l) + \bz, ~\bz \sim \mathcal{CN}\left(\mathbf{0},\sigma^2\mathbb{I}_N\right),
\label{eq:meas_model}
\end{equation}
where $g_l\in \mathbb{C}$ are the unknown complex gains. 
The signal to noise ratio for $l^{\text{th}}$ sinusoid is given by $\text{SNR}_l = |g_l|^2/\sigma^2$. The goal of the algorithm is to provide reliable estimates of $\{(g_l, \omega_l) : l = 1,2,\dots, K\}$ and $K$, the number of sinusoids in the mixture. 

The preceding model and its variants have many applications. For a linear array with $N$ elements with inter-element spacing $d$, the response 
corresponding to angle of arrival or departure $\theta$ relative to broadside is given by $\bx(\omega)$, where $\omega =  2\pi(d/\lambda)\sin(\theta)$ is the spatial frequency corresponding to $\theta$, and $\lambda$ denotes the carrier wavelength.  For estimation of a multipath channel $h(t) = \sum_{l=1}^K g_l \delta (t - \tau_l )$, the channel transfer function $H(f) = \sum_{l=1}^K g_l e^{- j 2 \pi f \tau_l}$.  Sampling uniformly in the frequency domain with spacing $\Delta f$ yields a mixture of sinusoids with $\omega_l = - 2 \pi \Delta f \tau_l$, reducing the problem of estimating delays to that of frequency estimation.  This directly models the operation of stepped frequency continuous wave (SFCW) for imaging a collection of point scatterers.  When the channel is ``seen through''
a pulse $p(t)$, as is often the case for channel estimation in communication applications, then the noiseless frequency domain signal is given by $Y(f) =\sum_{l=1}^K g_l P(f) e^{- j 2 \pi f \tau_l}$.  A simple extension of our algorithm to handle weighted sinusoids applies to this setting.
We do not provide detailed discussion here due to lack of space, but the code that we have made available \cite{online_code} does provide the required flexibility.

In this paper, we are interested in the setting where $K$ is a small integer, so that the underlying signal has a sparse representation in the atomic set of unit norm sinusoids. The goal is to find the best sparse approximation of $\by$ using atoms in the dictionary of sinusoids. A promising algorithm in sparse approximation theory is Orthogonal Matching Pursuit (OMP) \cite{Tropp2, Tropp}, a greedy algorithm that iteratively identifies the atom that yields the greatest improvement in approximation quality.  If the structure of the atomic set is ``simple,'' so that identifying
the ``best'' atom in each iteration needs a small amount of computation, then OMP becomes computationally attractive. Unfortunately, for our atomic set, searching over a continuum of atoms is not possible. An approximation that is often employed to overcome this problem is to discretize the set of frequencies, hoping that the signal still admits a sparse representation in this finite set. However, as discussed in \cite{mismatch}, no matter how finely we grid the parameter space, the true underlying atoms of $\by$ need not lie on the grid. This ``off-grid effect,'' or basis mismatch, degrades the performance of reconstruction algorithms significantly. The authors in \cite{jacques} propose inserting a gradient-based local search in a Matching Pursuit \cite{matching_pursuit} framework, as a means of alleviating the off-grid effect. In this paper, we go one step further by incorporating a Newton-based Cyclic Refinement step into the OMP framework: this not only
sidesteps the off-grid effect, but also enhances performance by refining previously estimated atoms in each iteration. 


\noindent
\subsection{Contributions}
\noindent Our key contributions are as follows:\\
(1) We propose Newtonized OMP (NOMP), which detects the best atom over a discrete grid, but avoids basis mismatch by adding a Newton refinement step, thus emulating pursuit over the continuum. In addition, we go beyond OMP by locally refining all estimated atoms in each iteration, thus re-evaluating estimates of previously detected sinusoids to incorporate the effect of the newly detected sinusoid. This second refinement step can be interpreted as feedback presented to the atoms we have already detected. We show that this feedback mechanism plays a crucial role in handling \textit{interference} among the underlying atoms of $\by$, 
yielding estimation accuracies far better than what would be possible by standard greedy pursuit algorithms (e.g., OMP and Matching Pursuit).
We do not require explicit estimates of model order, and provide a stopping criterion based on CFAR (constant false alarm rate) based on an estimate of the noise variance.\\
(2) We prove convergence of NOMP by providing upper bounds on the number of iterations. Moreover, we derive a  bound on the convergence rate,
showing that by choosing a fine enough grid for detection, the bound approaches the rate of convergence for OMP on the continuum.\\
(3) We show that the algorithm is near-optimal by numerical comparisons against the \CRB~(CRB) \cite{CRLB} in a variety of settings. When the frequencies of the sinusoids in the mixture are well-separated, NOMP is able to achieve the CRB for any sinusoid that has an SNR greater than a certain threshold. Moreover, NOMP is able to resolve closely-spaced frequencies with near-optimal accuracy, as long as there is enough disparity in the SNRs of different sinusoids. \\
(4) We analyze the computational complexity of NOMP.  Our numerical experiments show that its run time is significantly smaller than that of recently proposed 
state-of-the-art Atomic norm Soft Thresholding (AST) algorithm \cite{Recht1}, and is even smaller than that of the classical MUSIC algorithm \cite{MUSIC}.
We evaluate the run time of the algorithm in a variety of settings. 
A freely downloadable software package implementing the proposed algorithm can be found in \cite{online_code}.

\subsection{Related work}
Line spectral estimation is a fundamental problem in statistical signal processing.
Classical (and popular) subspace methods such as MUSIC and ESPRIT \cite{MUSIC, ESPRIT} exploit the low-rank structure of the autocorrelation matrix. 
One of the major advantages of these methods is the capability of resolving multiple closely-spaced frequencies at high SNR. Both MUSIC and ESPRIT have been shown to be asymptotically optimal in the limit of infinite SNR \cite{music_asym}, but their performance degrades at medium and low SNRs. 
 Another family of DFT-based classical methods \cite{Duda, Duda_survey}, typically have lower computational complexity and estimation accuracy similar to that of subspace methods \cite{Duda_survey, Vu}. 

More recent techniques using convex optimization cast the frequency estimation problem as that of finding a sparse approximation of the received signal using an infinite-dimensional dictionary of sinusoids.  It is shown in \cite{Candes1} that, in the absence of noise, total-variation norm is able to locate frequencies with infinite precision, as long as the minimum frequency separation exceeds $4\times\DDFT$ where $\DDFT = 2\pi/N$ is the Discrete Fourier Transform (DFT) grid spacing. The  sufficient condition on required minimum separation has been recently improved to $2.52\times\DDFT$ in \cite{Carlos}. An extension to noisy scenarios is provided in \cite{Candes2}.  Another approach is Atomic norm Soft Thresholding (AST) \cite{Recht1}, \cite{Tang}, which provides theoretical guarantees of noise robustness in terms of mean squared error (MSE). Both total-variation norm and atomic norm are generalizations of the $\ell_1$ norm to infinite-dimensional settings. Solving these optimization problems involves solving the Lagrange dual which takes the form of a semi-infinite program (SIP) with finite-dimensional decision variables and infinitely many constraints. For the sparse frequency estimation problem, \cite{Candes2} and \cite{Recht1} reformulate the dual as a semidefinite program (SDP),
which enables numerical optimization.
Similar reformulation for other problems seems to be difficult \cite{Tang}. A pragmatic approach is to use Lasso optimization on a highly oversampled grid as an approximation for AST \cite{Recht1, Tang}. Both AST and Lasso are benchmarks that we compare our proposed algorithm against in our numerical experiments. 



NOMP can be viewed as coordinate optimization over a continuum.  Coordinate-wise descent has been widely used for sparse
approximation; for example, such methods are competitive for solving Lasso type problems \cite{Tibshirani, TongTong}.
Preliminary results in \cite{Hassibi} show that coordinate descent to a relaxation of the AST problem can be a means to speed
up implementation.

There is a large body of work on feedback for improving the performance of iterative greedy algorithms for sparse approximation \cite{cosamp}, \cite{forward_backward}, \cite{IHT}, \cite{OMPR}, \cite{RELAX},  \cite{IAA}: information from recent iterations is used to remove errors introduced by previous steps. For example, \cite{forward_backward} introduces a forward-backward greedy algorithm that allows \mbox{\emph{aggressive}} backward steps (discarding small-magnitude atoms) after a greedy forward step. In contrast, NOMP, by virtue of its continuous parameterization, can employ a \emph{mild} form of feedback by locally refining the set of estimated atoms, thus allowing an atom to be replaced by a nearby, highly correlated atom, in the continuum. The idea of embedding a local refinement step in OMP has been proposed in \cite{BLOOMP} in a discrete setting, but this algorithm does not address basis mismatch,
and assumes that the model order is known {\it a priori.}

Many theoretical results on OMP \cite{Tropp, Greed, OMP_RIP} are based on assumptions such as incoherence or Restricted Isometry Property (RIP) for the underlying dictionary. Such approaches do not work for analyzing NOMP, since nearby atoms are highly correlated for continuous frequency estimation. Instead, our convergence analysis borrows tools from analysis of AST-based line spectral estimation \cite{Recht1}, along with observations following from our CFAR-based design.


For a single sinusoid, frequency estimation using coarse detection followed by Newton refinement was proposed three decades ago by Abatzoglou \cite{first_Newton}. This was recently adapted for estimation of a single delay in \cite{Bidigare2012}, and shown to approach
estimation-theoretic bounds.  In prior work involving some of the authors, we have used sequential algorithms similar to NOMP for millimeter wave spatial channel estimation with compressive measurements \cite{Dinesh_ITA12, Dinesh_Allerton12}. The NOMP algorithm in the present paper is an improvement on these, with a principled CFAR-based stopping criterion. 
We present NOMP within an application-independent abstraction, recognizing the fundamental nature and widespread utility of the frequency estimation problem.
We have reported on a version of NOMP in a recent conference
paper \cite{globalsip} (although the algorithm in \cite{globalsip} omits a least squares step included here for establishing convergence rate results, but with little impact on practical performance).  The present paper goes beyond \cite{globalsip} in providing a detailed convergence analysis, and a comprehensive comparison with the state of the art.

A closely related algorithm to NOMP is proposed in \cite{hansen}, which employs a Bayesian framework
for frequency estimation, using Newton refinements for updating the frequencies.  The details are different from our non-Bayesian, CFAR framework, and
convergence analysis is not provided, but the benefits of Newtonization are also evident in the numerical results in \cite{hansen}. 

\noindent {\bf Outline:} We present our algorithm in Section \ref{sec:algorithm}. In Section \ref{sec:CFAR}, we present the CFAR-based stopping criterion,
together with a simplified analytical characterization of false alarm and miss probabilities.
In Section \ref{sec:OSF}, we show why oversampling is essential for discretization followed by refinement to emulate pursuit over the continuum.
We discuss convergence in Section \ref{sec:convergence}.
In Section \ref{sec:sims}, we report on numerical experiments, and compare NOMP with other methods in terms of estimation accuracy and computational complexity. Section \ref{sec:extensions} discusses extension of the algorithm to more general settings, with illustrative numerical results for compressive measurements.
Section \ref{sec:conclusions} concludes the paper. \\ 
We set $N=256$ throughout in our numerical results (other values of $N$ yield entirely similar trends).\\
\noindent {\bf Notation:} Complex conjugate transpose of $v$ is denoted by $v^{H}$. $\Re\{a\}$ is the real part of complex number $a$. The Moore-Penrose pseudo-inverse of matrix $A$ is denoted by $A^{\dagger}$. The distance between any two frequencies $\{\omega_l, \omega_k\}$ is defined by $\text{dist}(\omega_k , \omega_l) \triangleq \min_{a\in\mathbb{Z}}|\omega_k - \omega_l + 2\pi a|$, i.e. the wrap-around distance when we restrict frequencies to lie in $[0,2\pi)$. The DFT matrix with unit norm columns and the corresponding grid spacing are denoted by $\mathcal{F}$ and $\DDFT$, respectively. The inner product between $v,u\in \mathbb{C}^{N}$ is defined as $\langle v , u\rangle = u^{H}v$.


\section{NOMP Algorithm} \label{sec:algorithm}

We first discuss estimation of a single sinusoid, and then build on it to generalize to a mixture of sinusoids.

\subsection{Single Frequency} We have $\by = g \bx(\omega) + \bz$. The Maximum Likelihood (ML) estimate of the gain and frequency are obtained by minimizing the residual power $\norm{\by - g\bx(\omega)}^2$, or equivalently, by maximizing the function
\begin{equation}
S(g, \omega) = 2\Re\{\by^{H}g\bx(\omega)\} - |g|^2 \norm{\bx(\omega)}^2.
\end{equation}
Directly optimizing $S(g, \omega)$ over all gains and frequencies is difficult. Therefore, we adopt a two stage procedure: (1) Detection stage, where we find a coarse estimate of $\omega$ by restricting it to a discrete set, (2) Refinement stage, in which we iteratively refine gain and frequency estimates.

For any given $\omega$, the gain that maximizes $S(g, \omega)$ is given by $\hat{g} = (\bx(\omega)^{H}\by )/\norm{\bx(\omega)}^2$. Substituting $\hat{g} $ in $S(g,\omega)$ yields that the generalized likelihood ratio test (GLRT) estimate of $\omega$ (treating $g$ as a nuisance parameter) is the solution to the following optimization problem:
\begin{equation}
\hat{\omega} = \arg \max_{\omega} ~G_{\by}(\omega),
\label{eq:glrt}
\end{equation}
where 
\begin{equation}
G_{\by}(\omega) = {\left.|\bx(\omega)^{H} \by|^2\middle / \left\Vert\bx(\omega)\right\Vert^2\right.}  
\end{equation}
is the GLRT cost function. We use this observation to find a coarse estimate of $(g, \omega)$ in the Detection stage.\\
\noindent{\bf Detection:} We obtain a coarse estimate of $\omega$ by restricting it to a finite discrete set denoted by $\Omega \triangleq \{k(2\pi/\gamma N): k = 0,1,\dots, (\gamma N -1)\}$, where $\gamma$ is the over-sampling factor relative to the DFT grid. For our simulation results, we set $\gamma = 4$. The outputs of this stage are $\omega_{\text{c}}\in \Omega$ that maximizes the GLRT cost function (\ref{eq:glrt}), and the
corresponding gain $(\bx(\omega_{\text{c}})^{H}\by) /\norm{\bx(\omega_{\text{c}})}^2$.\\
\noindent{\bf Refinement:} Since $\omega$ can take any value in interval $[0, 2\pi)$, we add a Newton-based refinement stage for estimation on the continuum. Let $(\ghat, \omegahat)$ denote the current estimate. The Newton step for frequency refinement is given by
\begin{equation}\label{eq:NewtonUpdate}
\hat{\omega}^\prime = \hat{\omega} - {\left.\dot{S}(\hat{g}, \hat{\omega}) /  \ddot{S}(\hat{g}, \hat{\omega})\right.}
\end{equation}
where
\begin{eqnarray}
\dot{S} (g,\omega) &=& \Re\{  (\by - g \bx(\omega))^{H}g (d\bx(\omega) / d\omega) \}\\
\ddot{S} (g,\omega) &=& \Re \{ (\by - g \bx(\omega))^{H}g (d^2\bx(\omega) / d\omega^2) \} \\\notag 
&&  \quad - |g|^2 \norm{d\bx(\omega) / d\omega}^2.
\end{eqnarray}
As we want to \emph{maximize} $S(g,\omega)$, we only apply the update rule \eqref{eq:NewtonUpdate} when the function is locally concave (i.e. \mbox{$\ddot{S}(\ghat, \omegahat) < 0$}). The gain parameter is then updated to maximize $S(g, \omegahat^\prime)$: $\hat{g}^\prime = (\bx(\omegahat^\prime)^{H}\by) /\norm{\bx(\omegahat^\prime)}^2$.\\
\noindent{\bf Refinement Acceptance Condition (\texttt{RAC}):} We \textit{accept} a refinement only if it leads to a strict improvement in $G_{\by}(\omega)$; that is, if $G_{\by}(\omegahat') > G_{\by}(\omegahat)$.
This ensures that an accepted refinement can only decrease the overall residual energy, and that the residual energy is non-increasing
throughout the course of the algorithm, which ensures convergence, as shown in Section \ref{sec:convergence}.

\subsection{Multiple Frequencies}
Let $\parSet = \left\{\left(g_l,\omega_l\right),~l=1,\dots,k\right\}$ denote a set of estimates of the parameters of the sinusoids in the mixture. Let 
\begin{equation}
\byres(\parSet) = \by -  \sum_{l=1}^{l=k}g_l \bx(\omega_l) 
\end{equation}
denote the {\it residual} measurement corresponding to this estimate. The following procedure is a direct generalization of the single sinusoid refinement algorithm to multiple frequencies. It proceeds by employing the single sinusoid algorithm to perform \emph{Newtonized coordinate descent} on the overall residual energy $\norm{\byres(\parSet)}^2$. One step of this coordinate descent involves cycling through all sinusoids in $\parSet$ in a predetermined order. In this process, suppose that we wish to refine the $l$-th sinusoid: we treat $\byres(\parSet\setminus\{(g_l, \omega_l)\})$ as our measurement $\by$ and employ the single frequency update step to refine $(g_l, \omega_l)$.

\begin{algorithm}
\caption{Newtonized Orthogonal Matching Pursuit\label{alg:NOMP}}
\begin{algorithmic}[1]

    \Function{extractSpectrum}{$\by$, $\tau$} 
   	\State $m\gets 0$, $\parSet_0 = \{\}$
	\vspace{0.1in}
	\While{$\max_{\omega \in \text{DFT}} G_{\byres(\mathcal{P}_{m})}(\omega)> \tau$}
	\vspace{0.2cm}
		\State $m \gets m + 1$ 
		\vspace{0.2cm}
		\State \textsc{Identify} \\ $\omegahat = \arg \max_{\omega \in \Omega} \left.G_{\byres(\mathcal{P}_{m-1})}(\omega)\right.$ \\ and its corresponding gain \\ $\ghat \gets \left.\left(\bx(\omegahat)^{H} \byres\left(\parSet_{m-1}\right) \right)\middle/\norm{ \bx(\omegahat) }^2\right.$ \label{step:identify}
		\State $\parSet'_{m} \gets \parSet_{m-1} \cup \{(\ghat,\omegahat)\}$
		\vspace{0.2cm}
		\State \textsc{Single Refinement}: Refine $(\ghat,\omegahat)$ using single frequency Newton update  algorithm ($R_{s}$ Newton steps) to obtain improved estimates $(\ghat^\prime, \omegahat^\prime)$. \label{step:single_refinement}
		\vspace{0.2cm}
		\State $\parSet''_{m} \gets \parSet_{m-1} \cup \{(\ghat^\prime,\omegahat^\prime)\}$
		\vspace{0.2cm}
		\State \textsc{Cyclic Refinement:} Refine parameters in $\parSet''_{m}$ one at a time: For each $(g, \omega)\in \parSet''_{m}$ we treat  \mbox{$\byres(\parSet''_{m} \setminus \{(g, \omega)\})$}  as the measurement $\by$, and apply single frequency Newton update algorithm. We perform $R_{c}$ rounds of cyclic refinements.
		Let $\parSet'''_m$ denote the new set of parameters. \label{step:cyclic_refinement}
		\vspace{0.2cm}
		\State \textsc{Update} all gains in $\parSet'''_{m}$ by least squares:\\
		$X \triangleq [\bx({\omega}_1) \dots \bx({\omega}_{m})]$, $\{\omega_l\}$ are the frequencies in $\parSet'''_m$\\
		$[g_1 \dots g_{m}]^{T} = X^{\dagger}\by$\\
		Let $\parSet_m$ denote the new set of parameters. \label{step:least_squares}
		\vspace{0.2cm}
	\EndWhile
	\vspace{0.1in}
	\State \Return{$\parSet_{m}$} 
	\EndFunction
  \end{algorithmic}
 \end{algorithm}

The NOMP procedure is summarized in Algorithm~\ref{alg:NOMP}. We now briefly discuss the role of its main components:\\
$\bullet$ \noindent \textsc{Single Refinement} (Step \ref{step:single_refinement}): Ideally, we want to identify $\omegahat$ which maximizes the GLRT cost function over the continuum. The \textsc{Single Refinement} step emulates search over the continuum by locally refining the estimate of $\omegahat$ obtained by picking the maximum over the discrete set $\Omega$. \\
$\bullet$  \textsc{Cyclic Refinement} (Step \ref{step:cyclic_refinement}) is where NOMP diverges from \emph{forward greedy methods} \cite{Barron} (in particular OMP) by providing \emph{feedback} for local refinements of previously detected sinusoids. This gives them an opportunity to better explain the received signal in light of new information regarding the presence of another sinusoid. This feedback is presented in the form of an updated residue. As we see in Sections~\ref{sub-sec:convergence-rates-empirical}~and~\ref{sec:sims}, this step is crucial for fast convergence and high estimation accuracy.\\
$\bullet$ \noindent \textsc{Update} by least squares (Step \ref{step:least_squares}): Here we update gains by projecting the received signal $\by$ onto the subspace spanned by the estimated frequencies. This ensures that the residual energy is the minimum possible for the current set of estimated frequencies. We see in Section~\ref{sec:convergence} that performing this projection step just prior to detecting a new sinusoid enables us to lower bound the convergence rate of NOMP by mirroring arguments used to establish bounds on OMP convergence  \cite{Barron}. 

We have left the number of refinement steps unspecified so far. For the simulations in this paper, we set $R_{s} = 1$ for every newly detected sinusoid in the \textsc{Single Refinement} step, and $R_{c} \in\{1,3,5\}$ refinement rounds in the \textsc{Cyclic Refinement} step, depending on the difficulty of the estimation problem. 

\subsection{Complexity analysis} 
\label{sec:complexity}
We analyze the computational complexity of the main steps of NOMP assuming that the algorithm runs for exactly $K$ iterations (i.e., perfect stopping). Checking whether the stopping criterion is satisfied is efficiently implemented using Fast Fourier Transform (FFT), with complexity $\mathcal{O}(KN\log(N))$. The \textsc{Identify} Step involves computing the GLRT cost function over the set of frequencies defined by $\Omega$. This can also be computed using FFTs in $\mathcal{O}(\gamma KN\log(\gamma N))$ time. The \textsc{Single Refinement} Step takes only $\mathcal{O}(R_{s}N)$ operations per sinusoid, hence the total cost for \textsc{Single Refinement} is $\mathcal{O}(R_{s}KN)$. The \textsc{Cyclic Refinement}  involves refining all  frequencies that have been estimated so far, and has overall complexity $\mathcal{O}(R_{c}R_{s}K^2N)$. If we directly compute the pseudo-inverse and apply it to the vector of observation (i.e., $X^{\dagger}\by$) in the \textsc{Update} Step, the complexity is $\mathcal{O}(NK^2 + K^3)$ per iteration, and the overall cost is $\mathcal{O}(NK^3 + K^4)$. However, we note that iterative methods (such as Richardson's iteration or conjugate gradient \cite{cosamp}) are extremely efficient in computing the least squares solution and can be used for speeding up the \textsc{Update} Step. Empirical observations suggest that the \textsc{Cyclic Refinement} step dominates the overall computational cost of NOMP.\\

 \noindent{\bf Minimum frequency separation:} When two frequencies, say $\omega_1$ and $\omega_2$, are ``very close'', intuitively, the mixture $g_1\bx(\omega_1) + g_2\bx(\omega_2)$ is explained ``very well'' by a single frequency, say as $(g_1+g_2)\bx(\omega_1)$.  Thus, a natural metric to characterize regimes for testing
algorithms for mixture frequency estimation is the \textit{minimum frequency separation} between any two sinusoids. We denote this by $\minSep = \min_{k\neq l} \text{dist}(\omega_k , \omega_l)$ and we would like our algorithms to work well even for small values of  $\minSep$. 

Without a minimum frequency separation condition, the estimation problem can be hopelessly ill-posed. This has been studied in detail in \cite{Candes1} using Slepian's work on prolate spheroidal sequences \cite{Slepian}. It has been shown that if the frequencies are clustered together, it becomes impossible for \textit{any} method to recover the information form the \textit{noisy} observations. It is important to note that in the limit of infinite SNR, however, one can still estimate a sparse clustered set of frequencies regardless of their separation (e.g., using Prony's method of polynomial interpolation \cite{Prony}).\\

\begin{figure}[htbp]
\centering
\includegraphics[scale=0.18] {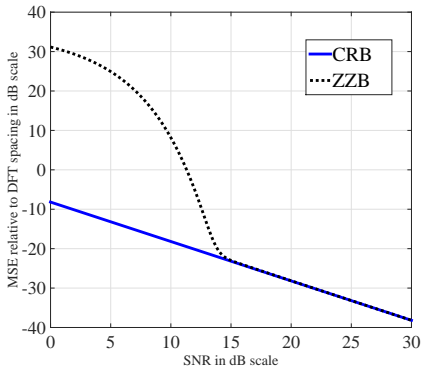}
  \caption{CRB and ZZB for estimating the frequency of a single sinusoid.}
  \label{fig:bounds} 
\end{figure}

\noindent{\bf Estimation Theoretic Bounds:}
Estimation theoretic quantities such as \CRB ~(CRB) and Ziv-Zakai Bound (ZZB) \cite{ZZB} provide lower bounds on the variance of estimators. For single frequency estimation, these bounds are given by \cite{Dinesh_TSP}
\begin{equation}\label{eq:crb}
\text{CRB}(SNR)  = {6\over SNR \times (N^2-1)},
\end{equation}
and
\begin{equation}\label{eq:zzb}
\text{ZZB}(SNR) = \int_{0}^{\pi} Q\left(\sqrt{SNR\left(1 - \left|{\sin(Nh/2)\over N\sin(h/2)}\right|\right)}\right) h~dh.
\end{equation}
In Figure \ref{fig:bounds}, we plot estimation error bounds as a function of $SNR = \left. \left\Vert\bx(\omega)\right\Vert^2 /\sigma^2\right. = \left. 1 \middle/\sigma^2\right.$. The ZZB has a distinct threshold behavior: large frequency estimation errors are inevitable in a ``low SNR'' regime below a
threshold, while the ZZB converges to the CRB when the SNR is large enough compared to this
threshold. In Section \ref{sec:sims}, we compare algorithms in terms of high SNR behavior relative to the CRB, and low SNR behavior relative to the ZZB threshold. 
Note that even if per sinusoid SNRs are higher than the ZZB threshold, the joint estimation problem for multiple sinusoids may still be ill-posed (e.g., if the frequency separation is too small).
\begin{remark}
We have defined ``integrated SNR'' obtained by dividing the total power of a sinusoid by the noise power per complex dimension, i.e., $SNR = {\mathbb{E}||\mathbf{x}(\omega)||^2\over \sigma^2}$. An alternative definition of signal to noise ratio is the ``per-sample SNR'', given by $SNR_{sample} = {\mathbb{E} [\mathbf{x}_{i}(\omega)]^2\over \mathbb{E}[\mathbf{z}_i]^2} = {1\over N} SNR$. For example, $SNR = 25$ dB (which is the nominal $SNR$ value in our simulations), corresponds to $SNR_{sample} \approx 1$ dB.
\end{remark}



\section{CFAR-based Stopping Criterion}
\label{sec:CFAR}
Detection problems face a tension between false alarm and missed detection (or miss, for short). In many detection problems, a model for the signal can be elusive. Thus, a common strategy is based on the Constant False Alarm Rate (CFAR) criterion \cite{CFAR1, CFAR2}, which only requires a model for noise. Here, we use the CFAR criterion to estimate model order (i.e. number of sinusoids in the mixture $K$): if the residual signal can be explained ``well enough'' by noise, up to a target false alarm rate, then we stop.
We show by simulation that the actual false alarm rate is close to the nominal being designed for.

We also estimate the probability of miss, taking into account the effect of noise but ignoring ``interference'' from other sinusoids.
The resulting receiver operating characteristic (ROC) turns out
to be in remarkable agreement with simulations.  This shows that, when the sinusoids are separated beyond a minimum separation, and when
their SNRs exceed the ZZB thresholds for individual sinusoids, then the probabilities of false alarm and miss are dominated by
noise rather than by inter-sinusoid interference.


\subsection{Stopping Criterion} 
The algorithm terminates when 
$$
G_{\byres(\parSet)}(\omega) = \left|\left<\byres(\parSet),\bx(\omega)\right>\right|^2 < \tau
$$ 
for all DFT frequencies $\{2\pi k/N~:~k=0,\dots,N-1\}$. In other words, we stop when 
$$
\norm{\mathcal{F}\byres(\parSet)}_{\infty}^2 < \tau,
$$
where $\mathcal{F}\mathbf{a}$ is the Discrete Fourier Transform of $\mathbf{a}$, and report $\parSet$ as our estimate of the sinusoids in the mixture. 

Suppose that we have already correctly detected all sinusoids in the mixture. In this case, the residual is $\byres(\parSet) \approx \bz$,
where $\bz  \sim \mathcal{CN}(\mathbf{0}, \sigma^2\mathbb{I}_N)$ (since $\mathcal{F}$ is a projection matrix, the statistics of
WGN are unchanged by it). It is easy to show that 
\begin{equation} \label{noise_ccdf}
\Pr\left\{\norm{\mathcal{F}\byres(\parSet) }_{\infty}^2 > \tau \right\} = 1 - \left(1-\exp(-\tau/\sigma^2)\right)^N.
\end{equation}
We choose our stopping criterion threshold $\tau$ so that $\Pr\left\{\norm{\mathcal{F}\byres(\parSet) }_{\infty}^2 > \tau\right\}  =  P_{\text{fa}}$, where $ P_{\text{fa}}$ is a nominal false alarm rate. Using (\ref{noise_ccdf}), we can explicitly compute this threshold as 
$$
\tau = - \sigma^2 \log\left({1-(1- P_{\text{fa}})^{1/N}}\right).
$$ 

A more easily interpreted expression can be obtained via asymptotics for large $N$ \cite{exponentials} (which provide excellent approximations for the moderate values of $N$ used in our numerical results).  Let $M_N \triangleq ||\mathcal{F}\byres(\parSet)||_{\infty}^2$. If $\byres(\parSet) \approx \bz$, we have $\mathbb{E}[M_N] = \sigma^2 \sum_{k=1}^{N}{1\over k} \approx \sigma^2 \log N$, and the asymptotic distribution of $E \triangleq M_N - \sigma^2\log N$ is given by $ \Pr\{E \leq x\} = \exp(-\exp(-x/\sigma^2))$. We set $\tau = \sigma^2 \log(N) + x$, for $x$ so that $\Pr\left\{E \geq x\right\}$ is equal to the nominal false alarm rate $P_{\text{fa}}$. This is given by $x =  -\sigma^2 \log\log\left( 1 \middle/ \left(1-P_{\text{fa}}\right) \right)$ and the resulting expression for the threshold $\tau$ is
\begin{equation}
\label{eq:p_fa}
\tau = \sigma^2\log(N) - \sigma^2 \log\log\left( 1 \middle/ \left(1-P_{\text{fa}}\right) \right).
\end{equation}

\begin{figure}[htbp]
\centering
\includegraphics[scale=0.19] {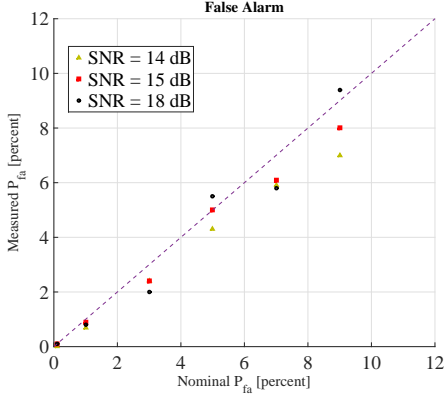}
  \caption{Nominal v.s. measured probability of false alarm.}
  \label{fig:pf_together} 
\end{figure}

Figure~\ref{fig:pf_together} plots measured versus nominal false alarm rates for different values of nominal $P_{\text{fa}}$. Each point in the plot is generated by $300$ runs of NOMP algorithm for estimating frequencies in a mixture of $K=16$ sinusoids of fixed nominal SNR. The minimum frequency separation $\minSep = 2.5\DDFT$. We declare a false alarm whenever NOMP overestimates the model order $K$. As shown in Figure \ref{fig:pf_together}, the empirical false alarm rate closely follows the nominal value at various SNRs.

\subsection{Probability of Miss} Define neighborhood $N_{\omega_i}$ around each true frequency $\omega_i$ by $N_{\omega_i} \triangleq \{\omega: \text{dist}(\omega, \omega_i) < 0.25\times\DDFT\}$. We declare successful detection of $\omega_i$ if at least one of the estimated frequencies lies in $N_{\omega_i}$, otherwise we declare a miss for $\omega_i$. A miss results from noise and inter-sinusoid interference,
but we only model noise here.  For the minimum separation considered here, we show using simulations that the empirical probability
of miss is only a little higher than the analytical estimate that we derive below.

\begin{figure}[htbp]
\centering
\includegraphics[scale=0.19] {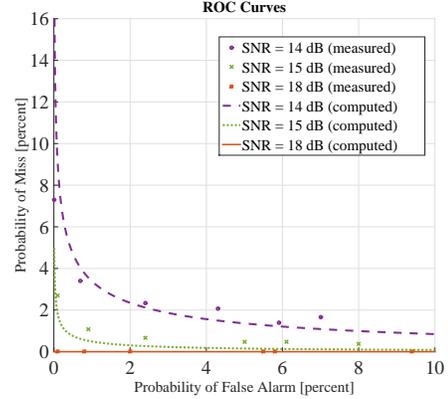}
  \caption{Measured v.s. computed ROCs}
  \label{fig:roc_together} 
\end{figure}

A sinusoid of amplitude $A$ leads to a maximum FFT value of $\alpha A$, where $\alpha \in [0,1]$ captures the amplitude reduction due to the grid mismatch. The magnitude of the maximum FFT coefficient, denoted by $M_{\text{fft}}$, is Rician with \mbox{$\text{Pr}\{M_{\text{fft}} < x\} = 1 - Q_{1} \left({\sqrt{2}\alpha A\over \sigma}  ,   {\sqrt{2} x\over  \sigma} \right)$}, where $Q_{1}$ is Marcum Q-function. The sinusoid is not detected by the algorithm if $M_{\text{fft}} < \sqrt{\tau}$, hence
\begin{equation}
P_{\text{miss}} = 1- Q_{1} \left(\alpha \sqrt{2SNR} ,  \sqrt{2\tau/\sigma^2} \right).
\end{equation}
Assuming uniform distribution for a frequency within a DFT grid interval gives us $\mathbb{E}[\alpha] = \mathbb{E}[\sin(N\omega/2)/(N\sin(\omega/2))] = 0.88$, where $\omega \sim \text{Uniform}[-\pi/N , \pi/N]$. Therefore, 
\begin{equation}
\label{eq:p_miss}
P_{\text{miss}} \approx 1- Q_{1} \left(0.88 \sqrt{2SNR} ,  \sqrt{2\tau/\sigma^2} \right).
\end{equation}
Putting  equations (\ref{eq:p_fa}) and (\ref{eq:p_miss}) together, we can characterize the ROC at various SNRs as shown in \mbox{Figure~\ref{fig:roc_together}}. The simulation parameters are the same as Figure \ref{fig:pf_together} .We see that for SNR$=18$ dB the probability of miss is negligible. When SNR goes below the ZZB threshold e.g. SNR$= 14$ dB, the probability of miss is bounded away from zero. This behavior is predicted by the ZZB threshold and our ROC analysis for single frequency estimation. However, as we see in Figure \ref{fig:roc_together}, they serve as excellent approximations for multiple frequency estimation. 

  
\begin{remark} 
For the simulations in this paper, we have used the CFAR-based stopping criteria for the NOMP algorithm.  However, a variety of other stopping rules, e.g., Bayesian Information Criteria (BIC) \cite{BIC} and Akaike Information Criteria (AIC) \cite{AIC}, can be easily adapted for use with the NOMP algorithm. In Section \ref{sec:model_order}, we investigate the performance of 
the BIC stopping rule (see Appendix I for a quick overview) as well as CFAR-based stopping criteria for NOMP.
\end{remark}

\section{The Need to Oversample}
\label{sec:OSF}
In this section, we show that oversampling is indeed required at the detection stage, in order for Newton refinements to converge to the maximum of the GLRT cost function. We ignore noise in these discussions.
The GLRT cost function is given by $G_{\by}(\omega) = |\sum_{l = 1}^{K} g_l h(\omega - \omega_l) |^2$, where $h(\omega)  = {\sin(N\omega/2)\over N\sin(\omega/2)}$ is the Dirichlet kernel. Characterizing the minimum required oversampling factor for this general setting is difficult because of the highly non-convex nature of $G_{\by}(\omega)$. Hence, we focus on a single frequency setting ($K = 1$). 

We wish to arrive at $\omega_1$ by optimizing $G_{\by}(\omega)$ using Newton refinements starting on a coarse grid. We note that $\arg \max_{\omega} G_{\by}(\omega)= \arg \max_{\omega} |g_1 h(\omega - \omega_1)|^2 = \arg \max_{\omega} |h(\omega - \omega_1)|^2$. 
We would like to characterize the minimum oversampling factor such that if we start off from the best guess of the maximum of $G_{\byres}(\omega)$ on the grid, the Newton refinement stage will take us to $\omega_1$.  That is, we must characterize how close to $\omega_1$ must the nearest grid point lie, so that Newton refinements will always take us to $\omega_1$ from this grid point.
Without loss of generality, we set $\omega_1 = 0$ (since we have shifted our frequency axis such that $\omega_1 = 0$, no grid point may lie on $0$).

We start by normalizing frequencies by the DFT spacing. In this scaled frequency axis, the Dirchelet kernel is given by $h(x) = {\sin(\pi x)\over N\sin(\pi x/N)}$, where $x  = \omega/\DDFT$ is sometimes referred to as the normalized frequency. As shown in \cite{Newton}, the Newton method converges to the solution of $h'(x) = 0$ quadratically if the initial guess $x_0$ lies in an interval $I$ around the true solution where the following conditions are met:
\begin{itemize}
\item $h''(x) \neq 0 \quad \forall x\in I$.
\item $h'''(x)$ is finite $\forall x\in I$.
\item $|x_0| < 1/M$ where $M \triangleq \sup_{x \in I} 0.5 \left|{h'''(x)\over h''(x)}\right|$.
\end{itemize}

\begin{figure}[htbp]
\centering
\includegraphics[scale=0.22] {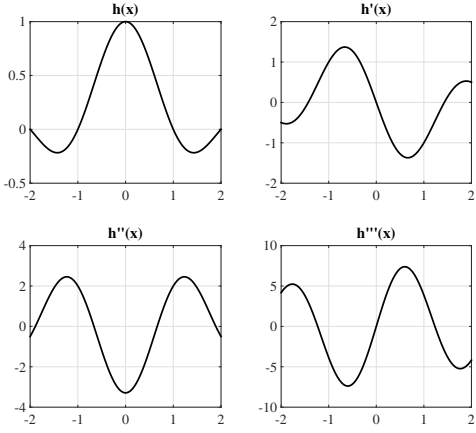}
  \caption{Dirichlet kernel h(x) and its derivatives.}
  \label{fig:Dirichlet} 
\end{figure}
Figure~\ref{fig:Dirichlet} shows the function $h(x)$ and its derivatives in a window around the origin. The first two conditions are met for any interval \mbox{$I \subset (-1, 1)$}. Therefore, it only remains to satisfy the third condition. By simple algebra one can see that if we set $I = (-0.45, +0.45)$, then we have $|x_0|<1/M$ for any \mbox{$x_0\in I$}. Therefore the maximum acceptable grid spacing is about $0.9$ which is equivalent to minimum oversampling factor $\approx 1.12$. This simple analysis shows that, even for a single sinusoid and no noise, we must sample beyond the DFT grid to ensure that the two-stage detection/refinement procedure successfully identifies the maximum of the GLRT cost function, thereby imitating pursuit over the continuum. Our simulations show that setting the oversampling factor at $\gamma = 4$ or more works very well, independent of the number of the sinusoids in the observations.



\section{Convergence}
\label{sec:convergence}
In this section, we first characterize convergence by providing upper bounds on the number of iterations of NOMP.
We then provide a bound on the rate of convergence of NOMP as a function of  the ``atomic norm'' of $\by$, and the oversampling factor $\gamma$. Our convergence results are pessimistic, in that they do not account for the effect of the refinement steps.  
We show by simulations the dramatic improvement due to refinements, by comparing NOMP to the following
variants of OMP:\\
{\bf Discretized OMP (DOMP)}:
This is standard OMP applied to the oversampled grid of sinusoids $\Omega$. We use our NOMP implementation with number of refinement steps set to zero, increasing the oversampling factor to $\gamma = 20$. Since DOMP can be interpreted as a special case of NOMP, the convergence analysis presented here is valid for DOMP as well.\\
{\bf NOMP without Cyclic Refinements \mbox{(NOMP--)}}: If we skip the \textsc{Cyclic Refinement} step of NOMP, then we get an algorithm that emulates OMP over the continuum of atoms. Note that NOMP-- does not have a feedback mechanism, hence it lies in the class of forward greedy methods. Our convergence analysis also holds here.


\subsection{Proof of Convergence:}
A trivial upper bound on the number of iterations of NOMP is the number of the observations $N$. This is a direct result of solving the least squares at Step \ref{step:least_squares} of the algorithm. After $N$ iterations, $X$ is a square full-rank matrix (no frequency is detected twice), hence the residue is zero for $(N+1)^{th}$ iteration and the algorithm terminates.

The following theorem states another upper bound on the number of iterations,
obtained by characterizing the amount by which the residual energy decreases when adding a new frequency to the set of estimated sinusoids.

\begin{theorem} The reduction of residual energy due to one iteration (adding a new sinusoid) is at least $\tau$. Consequently, $\min\{N, {\left.\norm{\by}^2/\tau\right.}\}$ is an upper bound on the number of iterations of the algorithm.

{\bf Proof:}
The residue at $m^{\text{th}}$ iteration of the algorithm is given by $\byres(\parSet_m) = \by - \sum_{l=1}^{l=m}g_l \bx(\omega_l)$. The energy of the residue in each iteration of the algorithm satisfies the following,
\begin{align}
\norm{\byres(\parSet_{m-1})}^2 &\stackrel{(a)}{=} \norm{\byres(\parSet'_m)}^2 + G_{\byres(\parSet_{m-1})}(\omegahat) \notag\\
&\stackrel{(b)}{\geq} \norm{\byres(\parSet'''_m)}^2 + G_{\byres(\parSet_{m-1})}(\omegahat) \notag\\
&\stackrel{(c)}{\geq} \norm{\byres(\parSet_m)}^2 + G_{\byres(\parSet_{m-1})}(\omegahat) \label{eq:reduction_proof} \\
&\stackrel{(d)}{\geq} \norm{\byres(\parSet_m)}^2 + \tau. \label{eq:reduction_tau}
\end{align}
where (a) follows from Step \ref{step:identify} of the algorithm where we project $\byres(\parSet_{m-1})$ orthogonal to the subspace spanned by $\bx(\omegahat)$ to get $\byres(\parSet_{m}')$. Inequalities in (b) follow from (\texttt{RAC}) checks performed whenever the single frequency refinement algorithm is invoked and (c) from the fact that least squares \textsc{Update} can only lead to a decrease in energy of the residual signal. (d) is a direct consequence of the stopping criteria of the algorithm. 

Inequality (\ref{eq:reduction_tau}) shows that the reduction of the residual energy due to detecting a new sinusoids is always greater than $\tau$. This result  bounds the number of iterations of the algorithm from above by $\left.\norm{\by}^2\middle/\tau\right.$, proving convergence. Combining this observation with the trivial upper-bound $N$ completes the proof. 
\end{theorem}

\subsection{Rate of Convergence:}
We first bound the maximum of the GLRT cost function $G_{\by}(\omega)$ over the  continuum of frequencies, in terms of that over the oversampled grid.
To this end, we borrow ideas used in \cite{Recht1} (Appendix C) for proving a similar property for the dual atomic norm.
We briefly introduce the notion of atomic norm (also known as dictionary norm) \cite{Barron} and specialize to the line spectral estimation problem.



The atomic set of unit norm sinusoids is given by \mbox{$\mathcal{A} =  \{e^{j\phi}\bx(\omega) : \phi, \omega \in [0,2\pi)\}$}. 
The atomic norm for $\by$ is defined by 
\begin{equation}
\norm{\by}_{\mathcal{A}} \triangleq \inf \left\{ t > 0 : \by \in t ~\text{conv}(\mathcal{A})\right\}.
\end{equation}
where $\text{conv}(\mathcal{A})$ denotes the convex hull of the points in $\mathcal{A}$. Since the centroid of the $\text{conv}(\mathcal{A})$ is at the origin, the atomic norm can be rewritten as \cite{Recht2, Banach}
\begin{equation}
\norm{\by}_{\mathcal{A}} \triangleq \inf \left\{\sum_{l} |g_l| : \by = \sum_{l}g_l \bx(\omega_l) , ~\bx(\omega_l)\in\mathcal{A}\right\}.
\end{equation}
Note that this is not the $\ell_1$ norm of $\by$, but the $\ell_1$ norm on the coefficients of the representation of $\by$ by elements of $\mathcal{A}.$ The atomic norm is typically small when its argument has a good sparse approximation \cite{Tropp2}.
The dual norm of $\norm{\cdot}_{\mathcal{A}}$ is defined by
\begin{equation}
\norm{\by}_{\mathcal{A}}^{*} = \sup_{\ba\in \mathcal{A}} \Re \{\langle \ba , \by\rangle\}.
\end{equation}
It is easy to see that
\begin{eqnarray}
\norm{\by}_{\mathcal{A}}^{*} &=& \sup_{\omega \in [0,2\pi)} \sup_{\phi \in [0,2\pi)} \Re \{ e^{i\phi} \langle \bx(\omega) , \by \rangle \} \notag\\
&=& \sup_{\omega \in [0,2\pi)} |\langle \bx(\omega) , \by\rangle| \notag\\
&=& \sup_{\omega \in [0,2\pi)} \sqrt{G_{\by}(\omega)}.
\end{eqnarray}
Directly borrowing from \cite{Recht1} gives us the following Theorem.\\
\begin{theorem}
\label{th:GLRTbounds}
\cite{Recht1} Maximizing the GLRT cost function (for the dictionary of unit norm sinusoids) over $[0,2\pi)$ is consistent with that over the oversampled grid $\Omega$ with oversampling factor $\gamma$. That is, we have
\begin{eqnarray}
\max_{\omega\in \Omega} \sqrt{G_{\by}(\omega)} &\leq& \sup_{\omega \in [0, 2\pi)} \sqrt{G_{\by}(\omega)} \\
&\leq& \left(1 - {2\pi \over \gamma}\right)^{-1} \max_{\omega \in \Omega} \sqrt{G_{\by}(\omega)}.
\end{eqnarray}
\noindent See (\cite{Recht1}, Appendix C) for a proof.
\end{theorem}

We need the following lemma to prove
Theorem \ref{th:rate_of_convergence}, which provides a pessimistic characterization of the convergence rate.

\begin{lemma}
\label{lemma:sequence}
Assume $\{a_n\}_{n\geq0}$ is a decreasing sequence of nonnegative numbers such that $a_0 \leq U$ and 
$$
a_n \leq a_{n-1}\left(1 - {a_{n-1}\over U}\right), \quad \forall n>0,
$$
then we have $a_n \leq {U\over n+1}$ for all $n\geq 0$. \\The proof is by induction \cite{Barron}. Suppose $a_{n-1}\leq {U\over n}.$ Either \mbox{$a_{n-1} \leq {U\over n+1}$}, in which case $a_n \leq {U\over n+1}$, or $a_{n-1} \geq {U\over n+1}$, in which case 
\begin{eqnarray}
a_n &\leq& a_{n-1}\left(1 - {a_{n-1}\over U}\right) \notag\\
&\leq& {U\over n}\left(1 - {{U\over n+1}\over U}\right) \notag\\
& = & {U\over n+1}. \notag
\end{eqnarray}
Hence, $a_n \leq{U\over n+1}$ for all $n \geq0$.
\end{lemma}

\begin{theorem}
\label{th:rate_of_convergence}
For all $\by$ such that $\norm{\by}_{\mathcal{A}} < \infty$, the residual energy of NOMP at the $\text{m}^{th}$ iteration satisfies
\begin{equation}
\norm{\byres(\mathcal{P}_m)} \leq (m+1)^{-1/2} \left(1 - {2\pi\over \gamma}\right)^{-1} \norm{\by}_{\mathcal{A}}.
\end{equation}

\noindent{\bf Proof:} From (\ref{eq:reduction_proof}), we have
\begin{equation}
\label{eq:th3_1}
\norm{\byres(\mathcal{P}_m)}^2 \leq \norm{\byres(\mathcal{P}_{m-1})}^2 - G_{\byres(\mathcal{P}_{m-1})}(\omegahat).
\end{equation}
$\byres(\mathcal{P}_{m-1})$ is the result of projecting $\by$ orthogonal to the subspace spanned by $\mathcal{P}_{m-1}$, therefore
\begin{eqnarray}
\norm{\byres(\mathcal{P}_{m-1})}^2 &=& \langle\byres(\mathcal{P}_{m-1}),\by \rangle\notag \\
&\stackrel{(a)}{\leq}& \!\! \norm{\by}_{\mathcal{A}} \norm{\byres(\mathcal{P}_{m-1})}_{\mathcal{A}}^{*}\notag\\
&=& \!\! \norm{\by}_{\mathcal{A}} \sup_{\omega \in [0,2\pi)} \sqrt{G_{\byres(\mathcal{P}_{m-1})}(\omega)}\notag\\
&\stackrel{(b)}{\leq}& \!\! \norm{\by}_{\mathcal{A}}\left(1 - {2\pi \over \gamma}\right)^{-1} \!\!\!\!\!\! \max_{\omega \in \Omega} \sqrt{G_{\byres(\mathcal{P}_{m-1})}(\omega)}, \notag
\end{eqnarray}
where $(a)$ follows by H\"{o}lder's inequality \cite{Recht3}, and $(b)$ is by Theorem \ref{th:GLRTbounds}. Let $\eta \triangleq \norm{\by}_{\mathcal{A}}\left(1 - {2\pi \over \gamma}\right)^{-1}$. From Step \ref{step:identify} of the algorithm we have 
$$
\omegahat = \arg \max_{\omega\in \Omega} \sqrt{G_{\byres(\mathcal{P}_{m-1})}(\omega)},
$$
hence,
\begin{eqnarray}
\norm{\byres(\mathcal{P}_{m-1})}^2  \leq \eta  \sqrt{G_{\byres(\mathcal{P}_{m-1})}(\omegahat)} \label{eq:th3_2}.
\end{eqnarray}
Combining (\ref{eq:th3_1}) and (\ref{eq:th3_2}), gives
\begin{eqnarray}
\norm{\byres(\mathcal{P}_m)}^2 \leq \norm{\byres(\mathcal{P}_{m-1})}^2 \left(  1 -   \eta^{-2}\norm{\byres(\mathcal{P}_{m-1})}^2 \right).
\end{eqnarray}
Using Lemma \ref{lemma:sequence} and the fact that
\begin{equation}
\norm{\byres(\mathcal{P}_{0})}^2 = \norm{\by}^2 \leq \norm{\by}_{\mathcal{A}}^2 \leq \eta^2,
\end{equation}
we have 
\begin{equation}
\norm{\byres(\mathcal{P}_m)}^2 \leq {\eta^2\over m+1}.
\end{equation}
In other words,
\begin{equation} \label{eq:NOMP_rate}
\norm{\byres(\mathcal{P}_m)} \leq (m+1)^{-1/2} \left(1 - {2\pi\over \gamma}\right)^{-1} \norm{\by}_{\mathcal{A}}.
\end{equation}
This proves the Theorem.
\end{theorem}

For the simulations we set $\gamma = 4$, but it is worth mentioning that, since we employ FFTs for detection over the oversampled grid, increasing $\gamma$ has marginal effect on the runtime of the algorithm. In fact setting $\gamma = 20$ leads to only about $5\%$ increase in runtime. 
If we compare (\ref{eq:NOMP_rate}) to the rate of convergence of OMP over the continuum given by \cite{Barron},
\begin{equation}
\norm{\byres(\mathcal{P}_m)}^2 \leq (m+1)^{-1} \norm{\by}_{\mathcal{A}}^2,
\end{equation}
we see that by choosing $\gamma$ large enough, the bound on the convergence rate specified in (\ref{eq:NOMP_rate}) approaches that of OMP over the continuum. Note that, in the derivation of (\ref{eq:NOMP_rate}) we have not considered the effect of the refinement steps, but imposing \texttt{RAC} ensures that refinements can only help speed up convergence. 

%
\subsection{Empirical rate of convergence:} \label{sub-sec:convergence-rates-empirical}
\begin{figure}[htbp]
\centering
\includegraphics[scale=0.19] {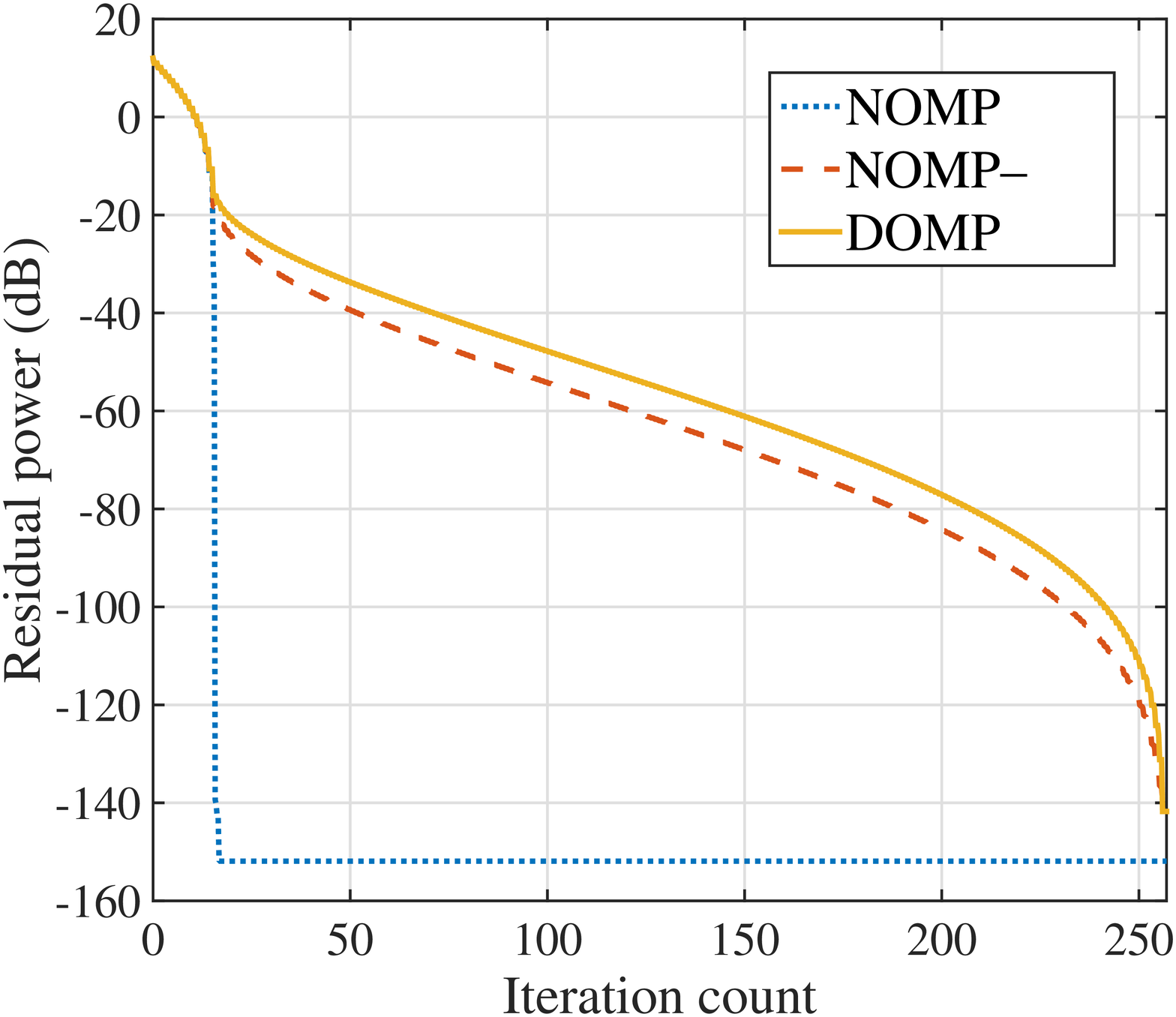}
  \caption{Convergence rates at noiseless case}
  \label{fig:convergence} 
\end{figure}

We use numerical simulations to highlight the convergence benefits of the refinement steps in NOMP compared to the bound specified in (\ref{eq:NOMP_rate}). We plot the mean residual energy (averaged over 1000 runs) as a function of the number of iterations in a noiseless setting. We set $K=16$, and $\minSep = 2.5\times \DDFT$. Figure \ref{fig:convergence} shows that light oversampling ($\gamma = 4$) followed by \textsc{Single Refinement} step (NOMP--) leads to a better convergence rate than having a large oversampling factor ($\gamma = 20$) and no refinements (DOMP). On the other hand, NOMP enjoys an extremely fast convergence rate due to the \textsc{Cyclic Refinement} step. In fact we see that for the setting where $\minSep$ is fairly large, the residual energy essentially drops to machine precision after $16$ iterations, which equals the number of sinusoids in the mixture.


\section{Simulation Results} \label{sec:sims}

Our performance measure is the mean squared error (MSE) of frequency estimation, and we compare the performance of NOMP against a number of benchmarks in various settings.\\
\noindent
{\bf Benchmarks:}  The MUSIC algorithm is implemented using a modified version of the \textsc{Matlab} routine \texttt{rootmusic}. The modifications are two-fold: (1) we use Minimum Description Length (MDL) criterion \cite{MDL} for estimating the number of sinusoids in the mixture; (2) MUSIC operates by constructing an estimate of the autocorrelation matrix of the observed data vector. To this end, we use a sliding window of size $W$, to generate multiple snapshots from the observation vector $\by\in\mathbb{C}^{N}$. The choice of $W$ has a significant impact on the performance of the algorithm, both in terms of estimation accuracy and run time: $W$ too large leads to an inaccurate estimate of the autocorrelation matrix and the signal subspace, whereas  too small a value effectively reduces the size of the observation window in time, degrading frequency estimation accuracy. We have found (empirically) that setting $W=96$ results in the best estimation accuracy for the nominal settings in our simulations. 


For sparse convex optimization, we consider  Atomic norm Soft Thresholding (AST) \cite{Recht1, Recht11}. We use the alternating direction method of multipliers (ADMM) \cite{ADMM} to implement AST, as suggested in \cite{Recht1}. The updates in ADMM are iterative (typically $100\sim 200$ iterations for each simulation run), and each iteration includes an eigenvalue thresholding step for an $(N+1) \times (N+1)$ matrix. This $\mathcal{O}(N^3)$ step dominates the computational cost of the ADMM method, and becomes very expensive for large $N$. 

The authors in \cite{Recht1} suggest solving Lasso as an alternative to the semi-definite program induced by AST. Lasso is solved on an oversampled frequency grid, using the 
highly optimized $\ell_2 - \ell_1$ software package  SpaRSA \cite{sparsa}. We set the tolerance parameter to be $10^{-3}$ (other than this, we use the default parameters): smaller values of the tolerance parameter (e.g. $10^{-4}$) increase runtime significantly, while providing marginal performance improvement. The regularization parameter in AST and Lasso formulation, suggested in \cite{Recht1}, is set to
$$
\text{reg} = \sigma \left(1 + {1\over \log N}\right)\sqrt{ \log N + \log(4\pi\log N)  }.
$$
The oversampling factor for the Lasso solver is set to $10$ in our simulations. While increasing the oversampling factor improves
Lasso performance, as mentioned in \cite{Recht1}, for an oversampled grid, the frequencies estimated by Lasso cluster around the true frequencies. In order to avoid drastically overestimating model order, we implement a simple \emph{clustering} scheme. In our simulations, we group frequencies into the largest number of clusters possible so that the frequency separation between any two sinusoids in different clusters is no smaller than $0.25 \times \DDFT$. After  clustering, we update the gains of the cluster centers by solving the least squares problem $\text{minimize}_{\{g_l\}} \norm{\by - \sum_{l} g_l \bx(\omega_l)}^2$. \\
\noindent
{\bf Newtonized Lasso (NLasso)}: We also compare the results of NOMP algorithm with an extension of the Lasso formulation. In this scheme, we first apply the Lasso solver to identify the frequencies over the highly oversampled grid. Then we run the \textsc{Cyclic Refinement} step of the NOMP algorithm in order to refine the estimated frequencies, in order to prevent error floors caused by the off-grid effect. The parameters of Lasso are unchanged and the number of refinement steps is set to $5$. Our simulations show that, for well-separated frequencies, the refinement step
significantly improves estimation accuracy while incurring a small increase in runtime compared to Lasso. However, the benefit of refinement for Lasso diminishes as we increase the difficulty of the estimation problem (small $\minSep$).\\
\noindent
{\bf Simulation set-up:}
We consider a mixture of $K = 16$ sinusoids of length $N = 256$. We perform $300$ simulation runs for each of the four scenarios characterized by $\minSep$ and SNR values. The settings for different scenarios are summarized in Table \ref{table:scenarios}.\\

\begin{table}[h]
\centering
\begin{tabular}{| c | c | c | c | c | c |}
\cline{1-3}
Scenarios &  \text{SNR (dB)}  & ${\minSep/\DDFT}$ \\  \cline{1-3}
1 & $\text{SNR}_{\text{nom}}$ & 2.5   \\ \cline{1-3}
 2 & $\text{SNR}_{\text{nom}}$ & 0.5  \\  \cline{1-3}
 3 & \text{Uniform}$[15,35]$ & 2.5  \\ \cline{1-3}
 4 & \text{Uniform}$[15,35]$ & 0.5  \\ \cline{1-3}
\end{tabular}
\caption{Settings of different Scenario}
\label{table:scenarios}
\end{table}

In Scenarios $1$ and $2$, the nominal SNR for each sinusoid is set as $\text{SNR}_{\text{nom}} = 25$ dB, whereas for Scenarios $3$ and $4$, the SNR values are chosen uniformly from $[15,35]$ dB, with mean equal to the nominal SNR of $25$ dB. 
%
%
In each simulation run, the gain magnitudes are set to $|g_l| = \sigma\sqrt{\text{SNR}_l}$, while the phases $\{\angle{g_l}\}$ are chosen uniformly from $[0, 2\pi)$. The frequencies are chosen uniformly at random from $[0, 2\pi)^K$ while respecting the minimum separation constraints specified by $\delomega$ (if the minimum separation criterion is not met, we sample again from $[0,2\pi)^K$). We plot the Complementary Cumulative Distribution Function (CCDF) of the squared frequency estimation error for all algorithms, along with the CRB (also a random variable, since it differs across realizations), and also compare against the DFT spacing, which is the resolution provided by coarse peak picking. See Appendix II for a quick overview of \CRB. The parameters of NOMP algorithm are set according to Table \ref{table:nomp_param} for different scenarios. 


\begin{table}[h]
\centering
\begin{tabular}{| c | c | c | c | c | c |}
\cline{1-5}
NOMP & Scenario 1  & Scenario 2 & Scenario 3 & Scenario 4 \\  \cline{1-5}
$R_c$ & 1 & 3 & 1 & 3  \\ \cline{1-5}
$R_s$ & 1 & 1 & 1 & 1  \\  \cline{1-5}
 $\gamma$ & 4 & 4 & 4 & 4  \\ \cline{1-5}
\end{tabular}
\caption{NOMP parameters at different scenarios.}
\label{table:nomp_param}
\end{table}

\begin{figure}[htbp]
\centering 
\subfigure[]{
\includegraphics[scale=0.18]{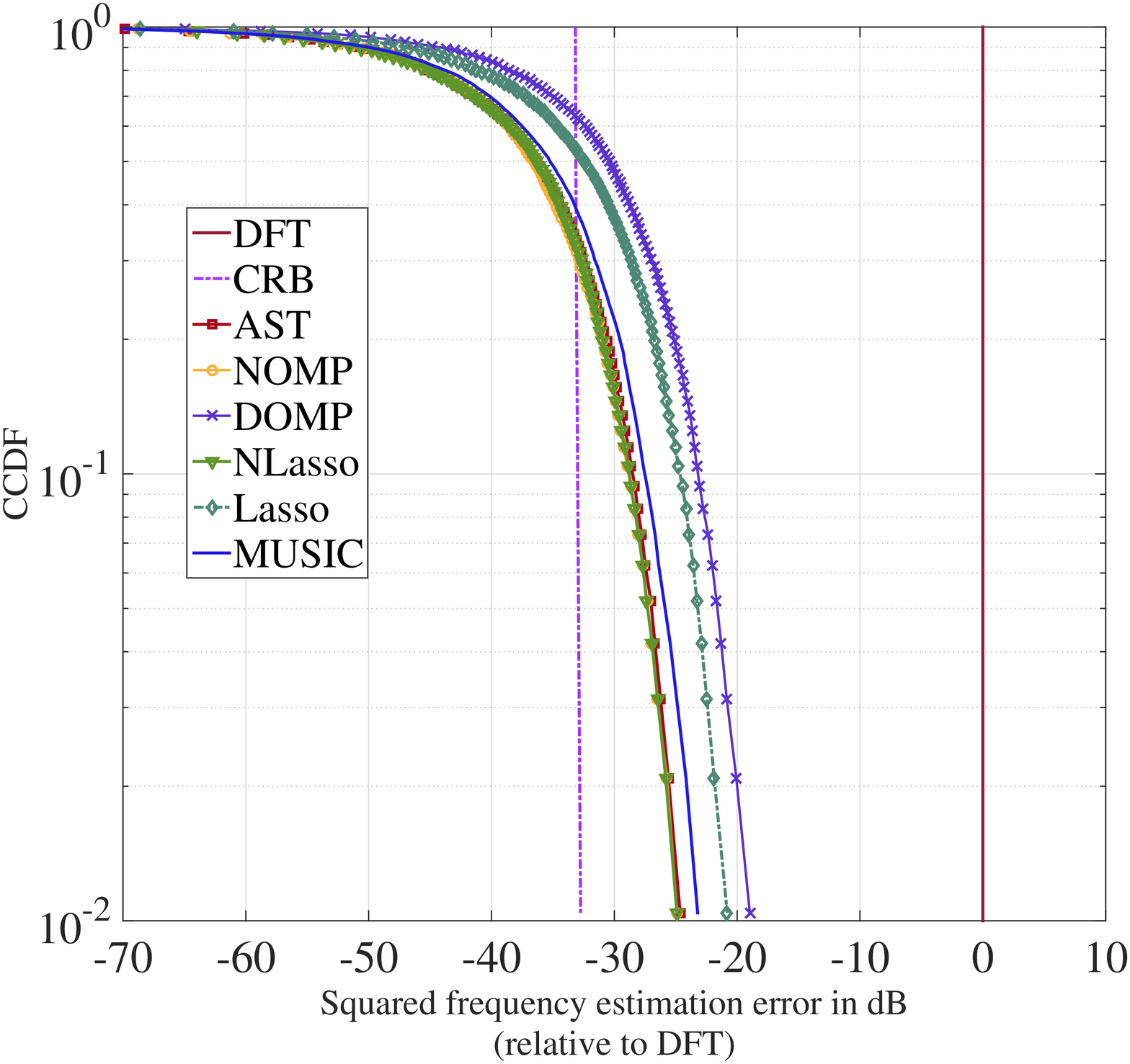}
}
\hspace{.2in} 
\subfigure[Zoomed in]{
\includegraphics[scale=0.18]{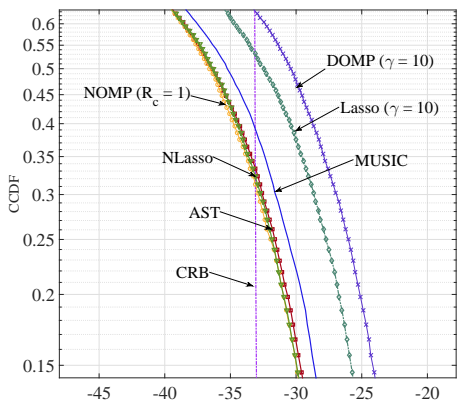}
}
  \caption{CCDF of the frequency MSE for Scenario 1.}
  \label{fig:scen1} 
\end{figure} 

\subsection{Frequency estimation accuracy}
\subsubsection{Distribution of error}

\begin{figure}[htbp]
\centering 
\subfigure[]{
\includegraphics[scale=0.18]{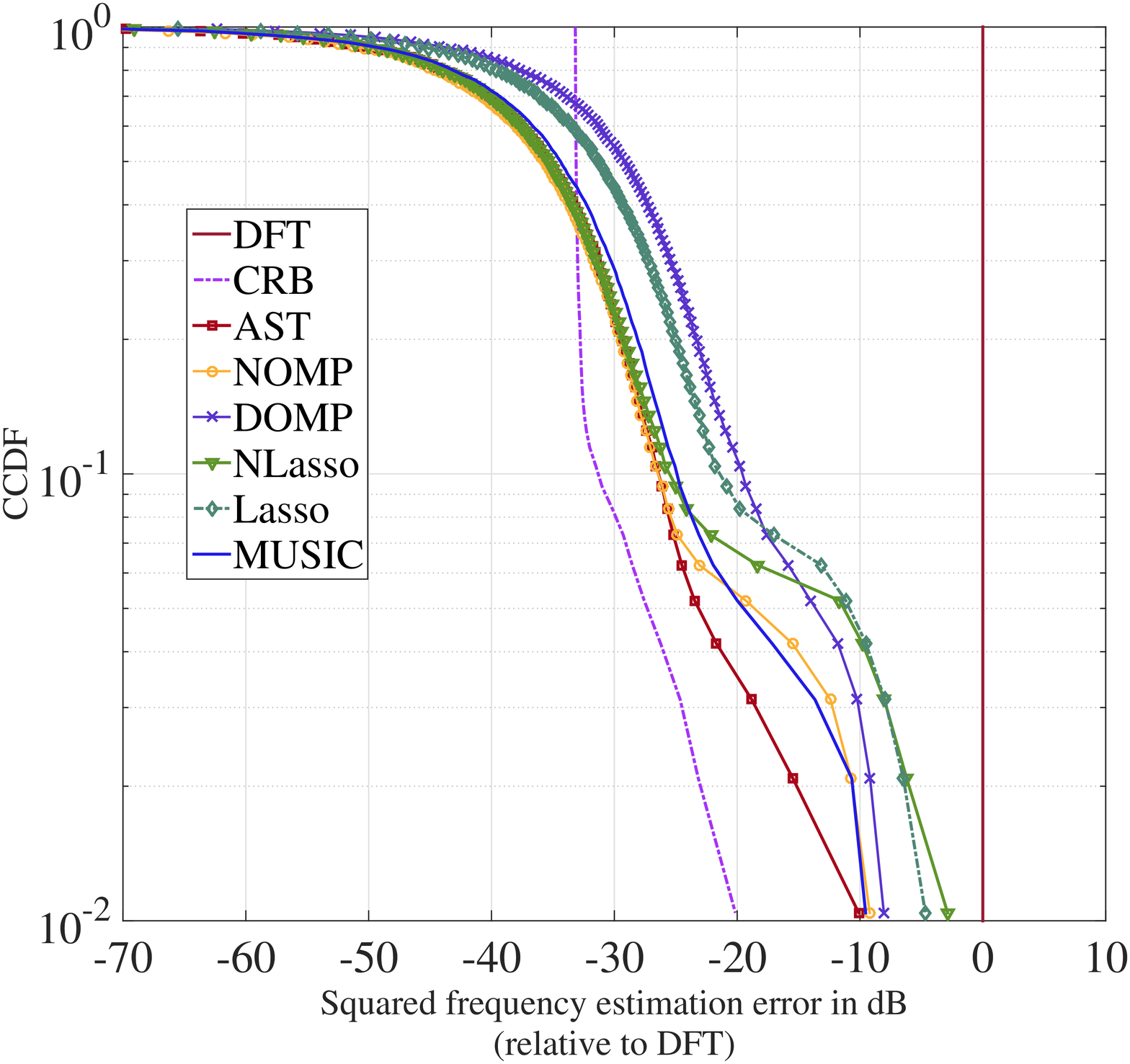}
}
\hspace{.2in} 
\subfigure[Zoomed in]{
\includegraphics[scale=0.18]{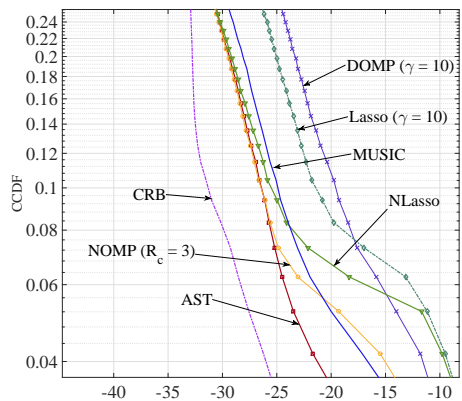}
}
  \caption{CCDF of the frequency MSE for Scenario 2.}
  \label{fig:scen2} 
\end{figure}

Let us first examine the CCDF of squared frequency estimation error in each scenario. Figure \ref{fig:scen1} shows that  NOMP, AST and NLasso  lead to very similar error distributions in Scenario~1, while outperforming the other methods. Unlike Lasso and DOMP, which suffer from the off-grid effect, MUSIC picks frequencies over the continuum, and achieves better estimation accuracy. Another observation is that if the frequencies are well separated (as in Scenario 1), then adding a refinement stage at the output of Lasso leads to a significant improvement in estimation accuracy. As we move to more difficult Scenarios, however, the performance of NLasso degrades compared to NOMP and AST as shown in Figure~\ref{fig:scen2} for Scenario 2. The reason is that the refinement stage of NLasso is able to improve the estimation accuracy only when the initial estimates provided by Lasso are close to the true frequencies. When Lasso fails in providing good initial estimates, there is no benefit in locally refining the frequencies. 

\begin{figure}[htbp]
\centering 
\subfigure[]{
\includegraphics[scale=0.18]{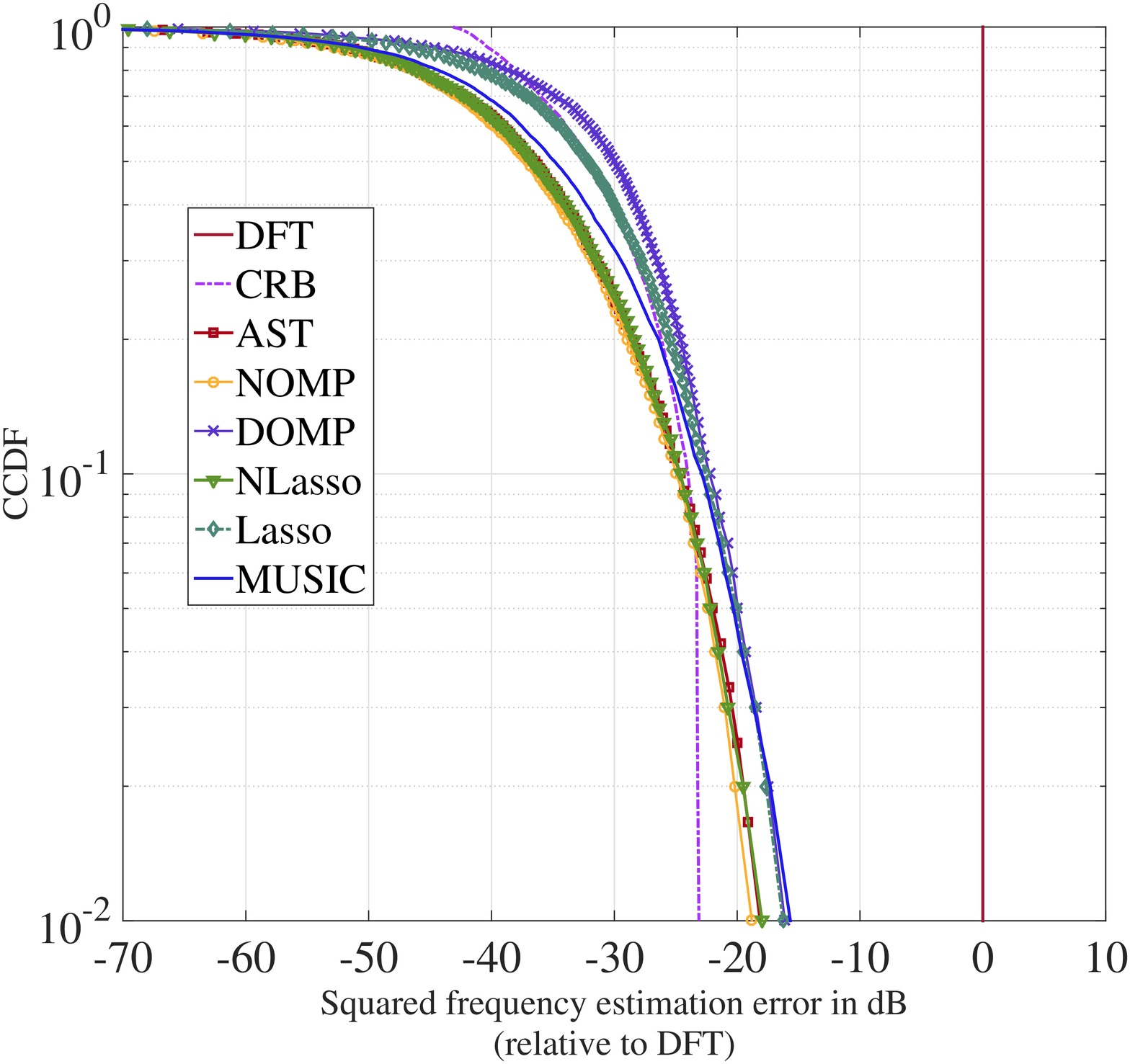}
}
\hspace{.2in} 
\subfigure[Zoomed in]{
\includegraphics[scale=0.18]{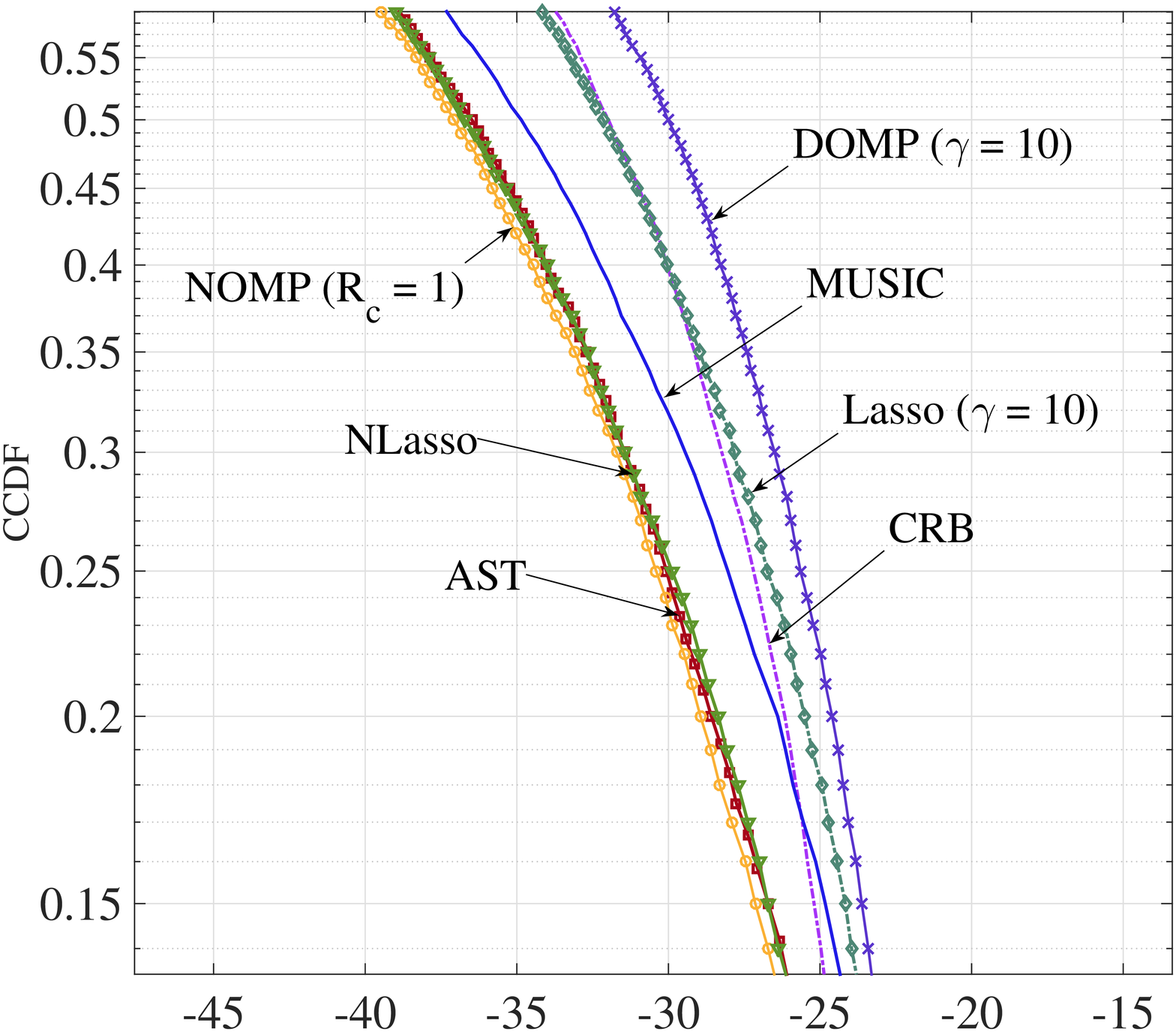}
}
  \caption{CCDF of the frequency MSE for Scenario 3.}
  \label{fig:scen3} 
\end{figure}

\begin{figure}[htbp]
\centering 
\subfigure[]{
\includegraphics[scale=0.18]{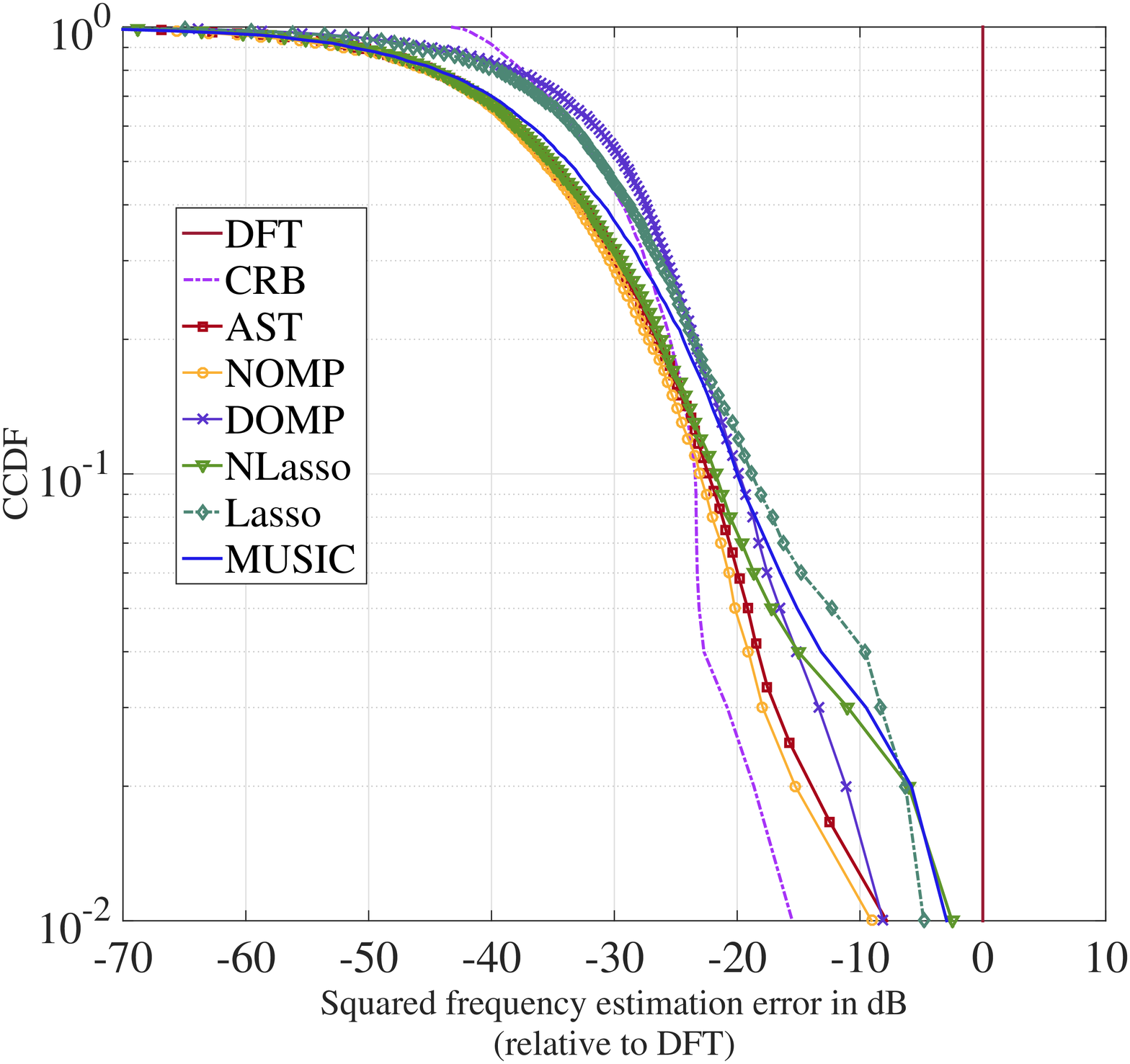}
}
\hspace{.2in} 
\subfigure[Zoomed in]{
\includegraphics[scale=0.18]{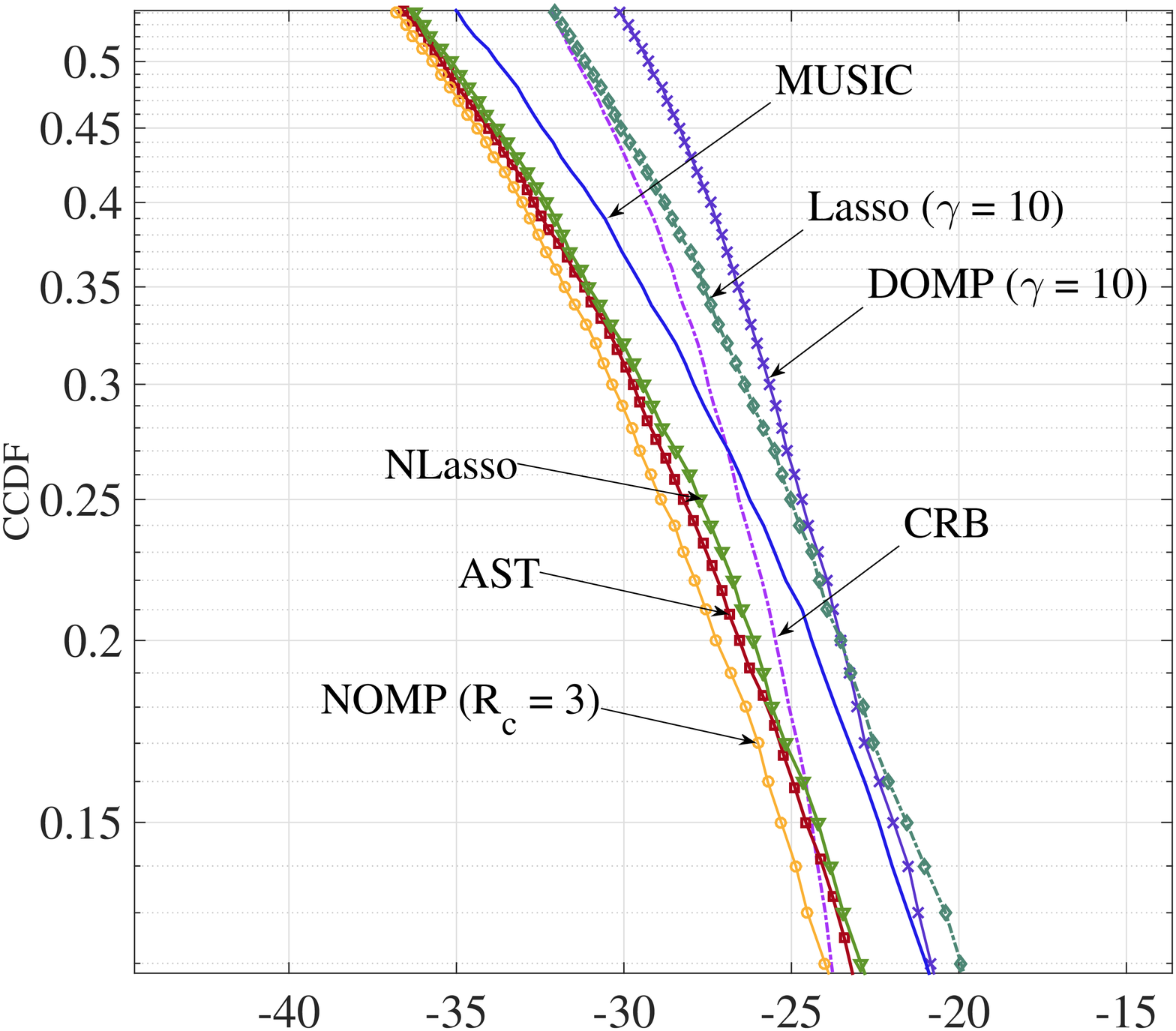}
}
  \caption{CCDF of the frequency MSE for Scenario 4.}
  \label{fig:scen4} 
\end{figure}

Figure \ref{fig:scen3} shows the distribution of error in Scenario 3. We see that the overall gap in the performance of different algorithms is reduced compared to the first two scenarios, with AST, NOMP and NLasso still achieving the highest estimation accuracy. In Scenario 4, NOMP achieves superior performance compared to all of the other methods, as shown in Figure \ref{fig:scen4}. 


\subsubsection{Mean squared error}
Here we examine the performance of different algorithms in terms of frequency estimation accuracy by looking at the normalized Mean Squared Error (MSE), defined by $\mathbb{E}[(\omega_{true} - \omega_{est})^2]/\DDFT^2$, in different scenarios. In Scenarios 1 and 3, where frequencies are well-separated, 
one hopes to get to the estimation accuracy of a single sinusoid (as if the other sinusoids did not exist).  We therefore use
the CRB and ZZB corresponding to a \textit{single sinusoid}, computed by (\ref{eq:crb}) and (\ref{eq:zzb}), respectively, as measures of optimality. 
In Scenarios 2 and 4, where frequencies can get close to one another, we compute the CRB empirically for each realization of the problem, and employ the mean CRB as a measure of optimality. 

\begin{figure}[htbp]
\centering 
\subfigure[Scenario 1]{
\includegraphics[scale=0.18]{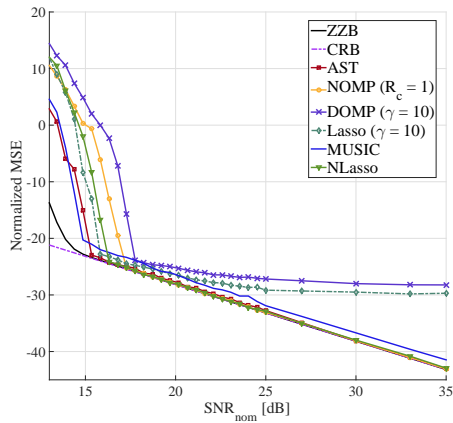}
}
\hspace{.2in} 
\subfigure[Scenario 2]{
\includegraphics[scale=0.18]{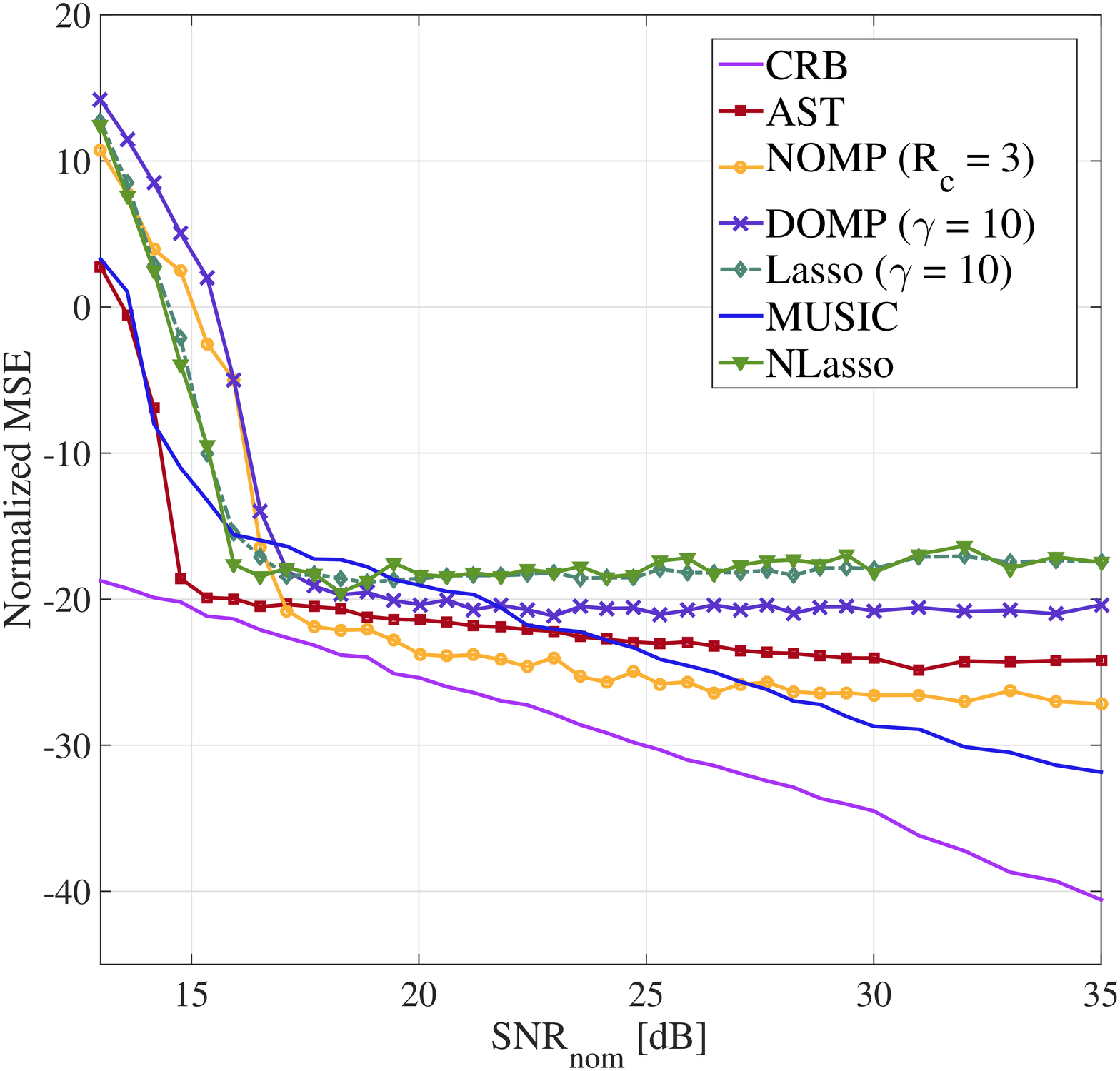}
}
  \caption{Normalized frequency MSE for Scenarios 1 and 2.}
  \label{fig:snr_scen12} 
\end{figure}

Figure \ref{fig:snr_scen12} shows the MSE of frequency estimation in Scenarios 1 and 2, for $\text{SNR}_{\text{nom}}$ taking values from $13$ dB to $35$ dB. In Scenario 1, if the nominal SNR is high enough, AST, NOMP, and NLasso all achieve the CRB. As we decrease the nominal SNR, all of the algorithms exhibit threshold behavior, well predicted by the ZZB threshold. The threshold SNR of AST is lower than that of other methods, showing its noise resilience.
MUSIC does not achieve the CRB, but closely follows the bound for all SNR values in this scenario. DOMP and Lasso on the other hand, reach performance floors. We examine this floor more closely in Section \ref{sec:asymptotic} to see whether it is a fundamental algorithmic limitation, or happens due to the off-grid effect. Figure \ref{fig:snr_scen12}-b corresponds to Scenario 2, where the separation between frequencies can be very small. We see that MUSIC is the only algorithm that benefits from increasing the nominal SNR. This goes back to the asymptotic optimality of MUSIC: for $SNR\to\infty$ and $K\ll N$, MUSIC is able to precisely determine the frequencies in the mixture, regardless of the separation between them \cite{music_asym}.

\begin{figure}[htbp]
\centering 
\subfigure[Scenario 3]{
\includegraphics[scale=0.18]{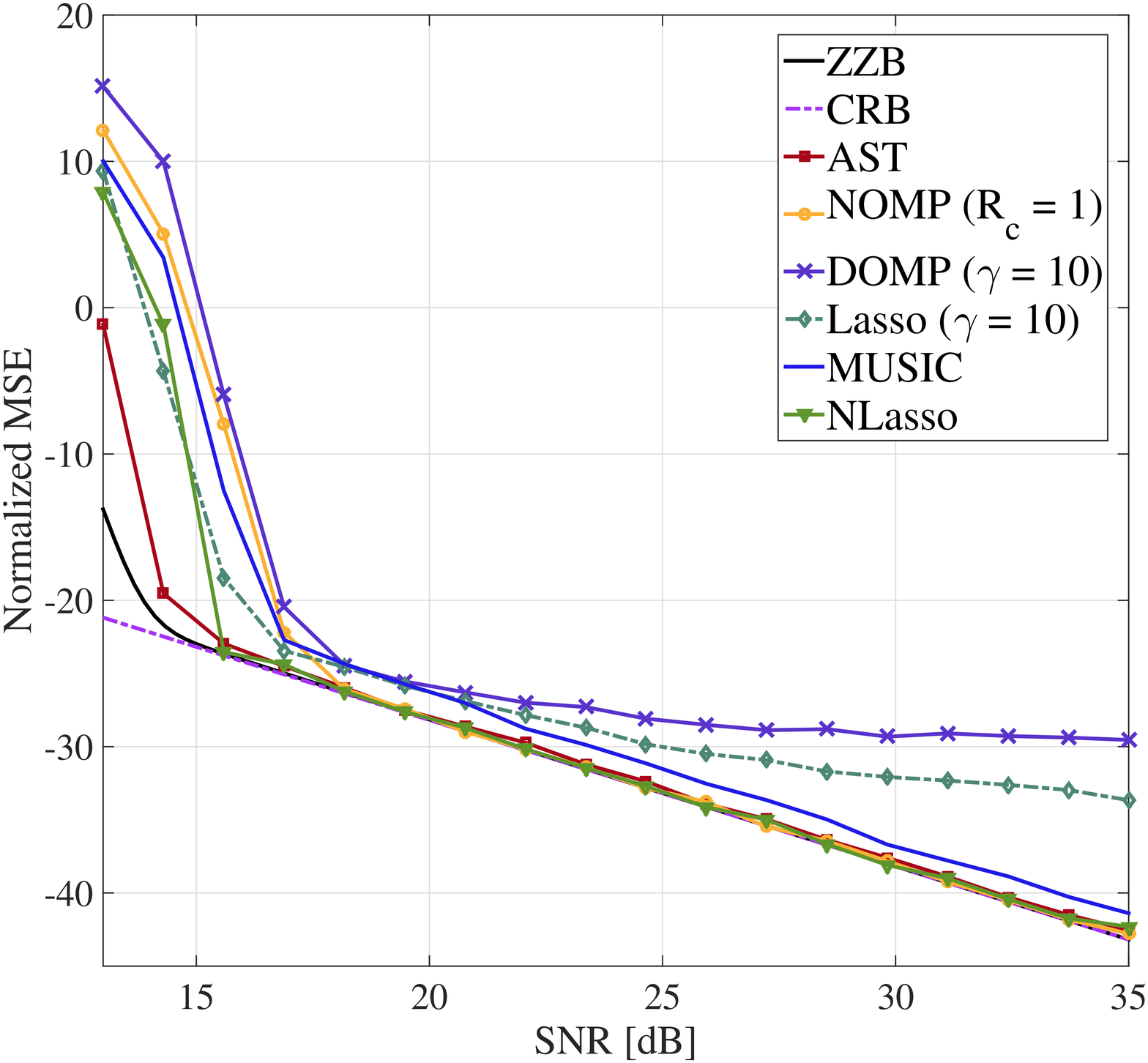}
}
\hspace{.2in} 
\subfigure[Scenario 4]{
\includegraphics[scale=0.18]{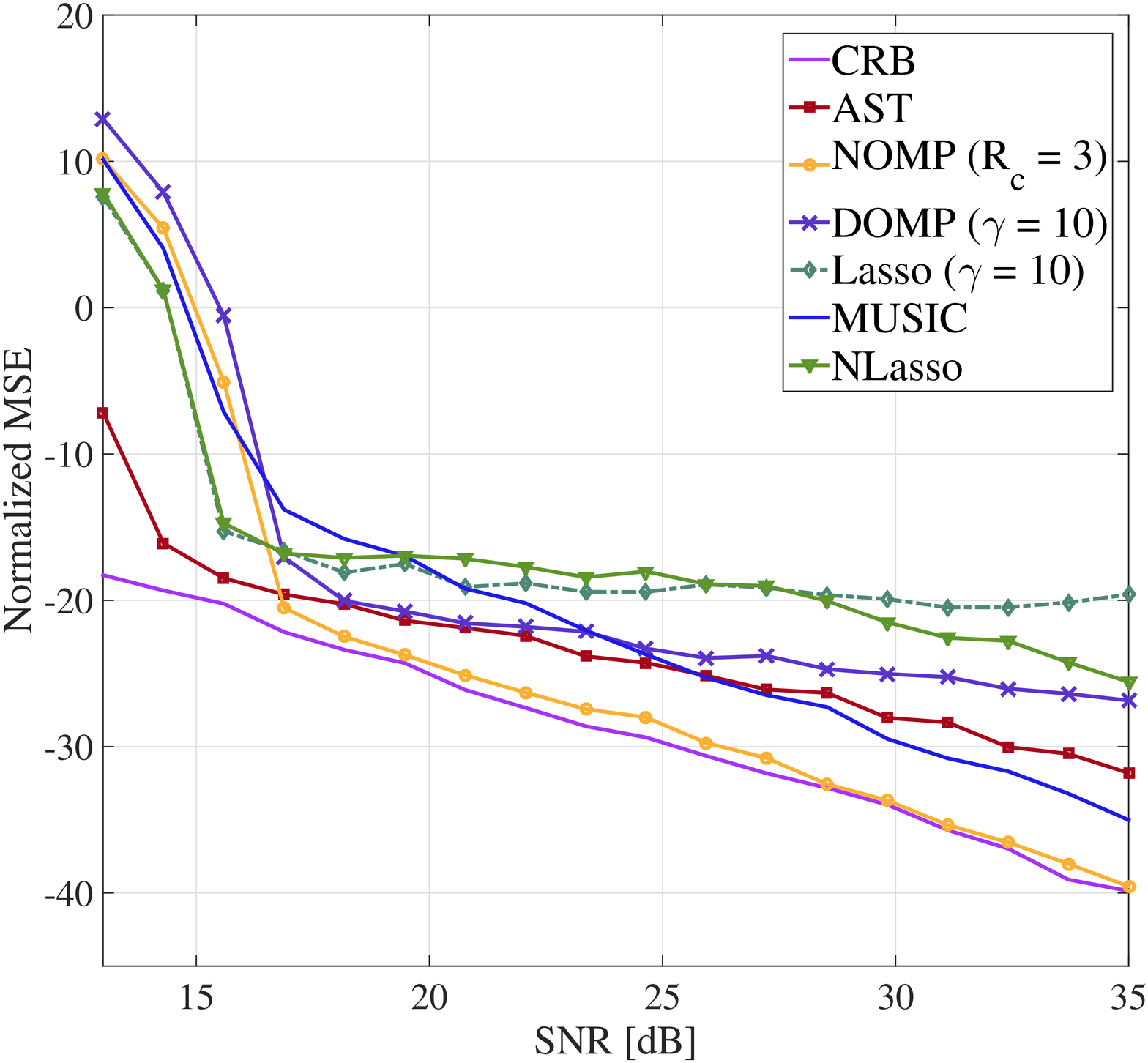}
}
  \caption{Normalized frequency MSE for Scenarios 3 and 4.}
  \label{fig:snr_scen34} 
\end{figure}

In Scenarios 3 and 4, the SNRs are drawn independently and uniformly at random from the interval $[15, 35]$ dB. In order to evaluate the frequency MSE in these scenarios, we fix the SNR of one of the sinusoids in the mixture at a given value, while letting the other $(K-1)$ SNRs to be realized randomly. Figure \ref{fig:snr_scen34}-a shows the frequency MSE curves corresponding to Scenario 3. As we expected from error distribution plots in Figure \ref{fig:scen3}, the gap in the performance of different algorithms has decreased compared to the first scenario. 
Figure \ref{fig:snr_scen34}-b corresponds to Scenario 4, in which NOMP outperforms all of the other algorithms, and tightly follows the CRB. This indicates that NOMP is highly successful in exploiting the disparity in SNRs across sinusoids in the mixture, in order to estimate closely-spaced frequencies. AST achieves the best performance at very low SNRs; however, its MSE curve stays bounded away from the CRB, with an expanding gap as we increase SNR. 

\subsubsection{Number of cycles of Newton refinement ($R_{c}$)} 
\label{sec:R_c_effect}
\begin{figure}[htbp]
\centering 
\subfigure[Scenario 2]{
\includegraphics[scale=0.18]{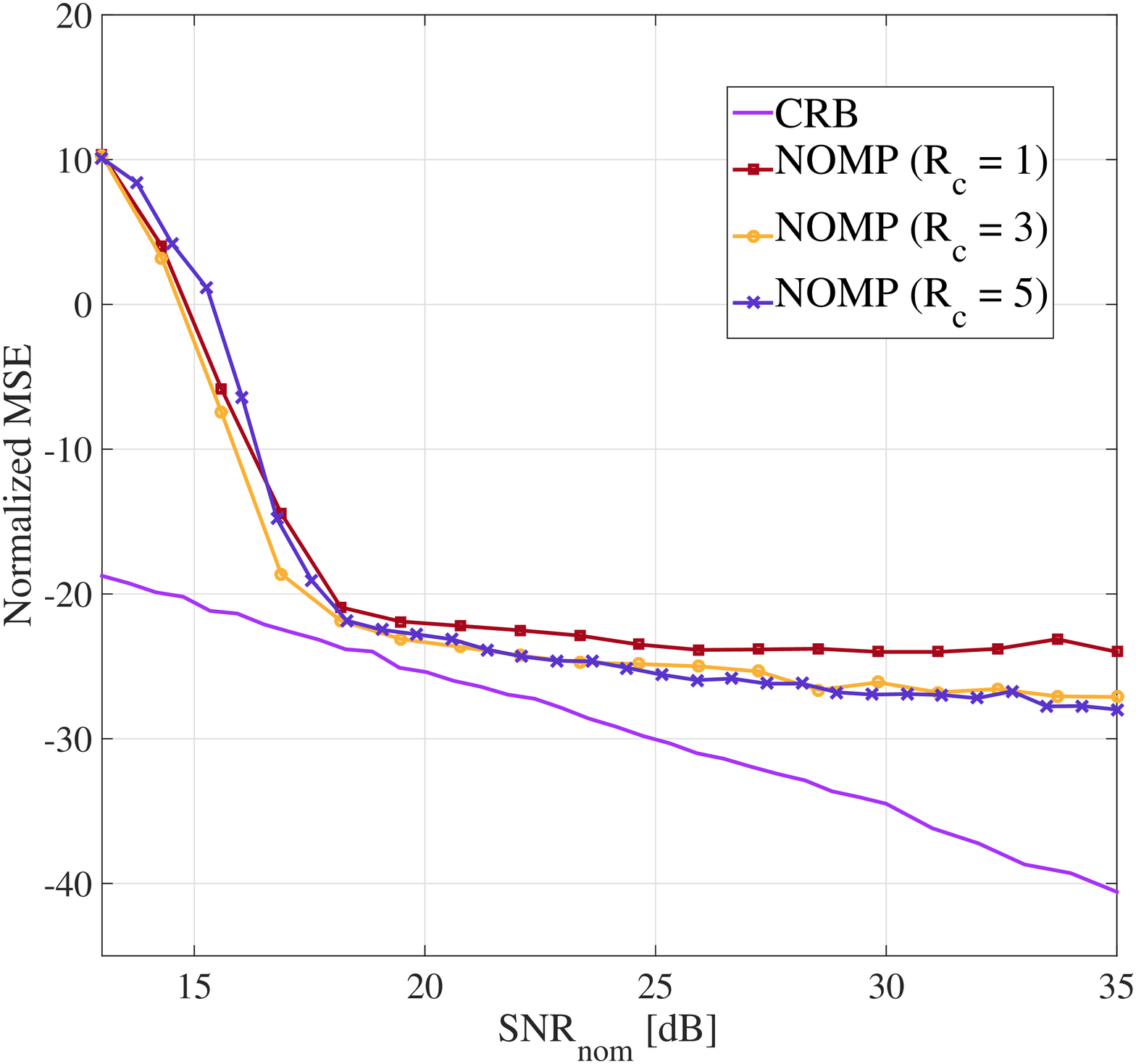}
}
\hspace{.2in} 
\subfigure[Scenario 4]{
\includegraphics[scale=0.18]{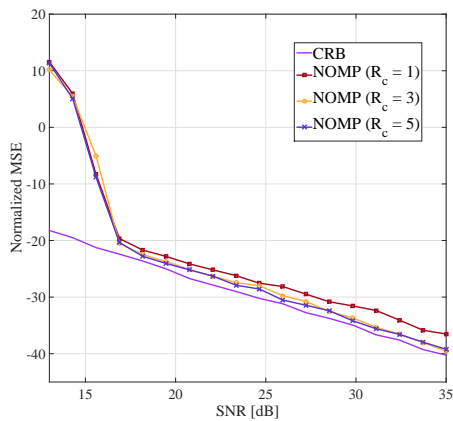}
}
  \caption{Performance improvement of NOMP with increasing the number of cyclic refinements in Scenarios 2 and 4.}
  \label{fig:nomp_scen24} 
\end{figure}

We have seen that NOMP is able to achieve the CRB in Scenarios 1 and 3, with one cycle of refining the sinusoids in each iteration, i.e., $R_c = 1$.   In this subsection, we want to highlight the effect of increasing $R_c$ in improving the frequency estimation accuracy of NOMP in Scenarios 2 and 4 where $\minSep = 0.5\times\DDFT$. Figure \ref{fig:nomp_scen24} shows the frequency MSE of NOMP for $R_c\in\{1,3,5\}$. We see that NOMP enjoys the benefits of having more rounds of refinement, but exhibits diminishing return as $R_c$ increases. In particular, increasing $R_c$ beyond $3$ cycles gives marginal improvement in estimation accuracy. 

\subsection{Computation time}
Table~\ref{table:time} summarizes the time needed for running $300$ simulations for each of the algorithms in different scenarios. We see that  DOMP is extremely fast at the expense of estimation accuracy. NOMP is faster than all of the other methods in all four Scenarios, while achieving remarkable frequency estimation accuracy. As we discussed in Section \ref{sec:complexity}, the \textsc{Cyclic Refinement} step has the complexity $\mathcal{O}(R_{c}R_{s}K^2N)$, and dominates the computational cost of NOMP. Table~\ref{table:nomp_time} shows the time needed for $300$ simulation runs of NOMP for different  values of $R_c$. Note that as we increase the difficulty of the estimation scenario, AST, Lasso and NLasso tend to take more time, while MUSIC and NOMP are unaffected.

\begin{table}[h]
\centering
\begin{tabular}{| l | c | c | c | c | c | c |}
\cline{1-7}
 Time [sec] &  \text{NOMP}  & \text{AST} &  \text{NLasso}  & \text{Lasso} & \text{DOMP} & \text{MUSIC}    \\  \cline{1-7}
 Scenario 1 & \text{  6.92} &1.09e3 & 29.68 & 26.63 & 2.66 & 19.83     \\ \cline{1-7}
 Scenario 2 & \text{  14.26} & 1.02e3 & 29.63 & 27.62 & 2.81 & 20.15  \\ \cline{1-7}
Scenario 3 &  \text{  6.88} & 1.12e3 & 34.32 &  32.06 & 2.79 & 20.07 \\ \cline{1-7}
Scenario 4 &  \text{  14.19} & 1.18e3 & 36.37 &  33.71 & 2.80 & 19.69  \\ \cline{1-7}
\end{tabular}
\caption{Time [sec] for 300 runs of each algorithm. Parameters of NOMP are set by Table \ref{table:nomp_param}}
\label{table:time}
\end{table}

\begin{table}[h]
\centering
\begin{tabular}{| l | c | c | c | c | }
\cline{1-4}
 Time [sec] &  $R_c = 1$ & $R_c = 3$ & $R_c = 5$   \\  \cline{1-4}
 Scenario 1 & \text{ 6.92} & 14.24 &  21.67 \\ \cline{1-4}
 Scenario 2 & \text{  7.01} & 14.26 &   21.71  \\ \cline{1-4}
Scenario 3 &  \text{   6.88} &  14.21 & 21.61 \\ \cline{1-4}
Scenario 4 &  \text{  6.96} & 14.19 & 21.50  \\ \cline{1-4}
\end{tabular}
\caption{Time [sec] for 300 runs of NOMP algorithm, for different numbers of cyclic refinements $R_c$.}
\label{table:nomp_time}
\end{table}

\subsection{Asymptotic regime} \label{sec:asymptotic}
\begin{figure}[htbp]
\centering
\includegraphics[scale=0.215] {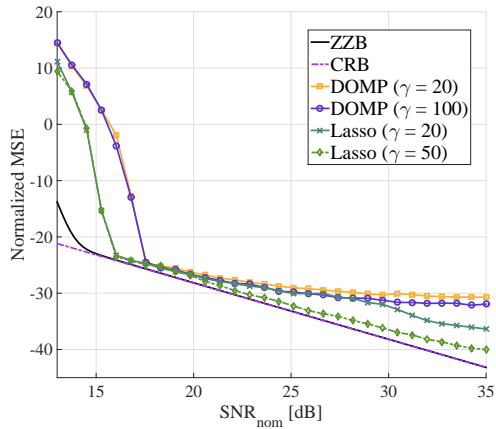}
  \caption{Frequency MSE for  Scenario 1 and highly over-sampled grid for Lasso and DOMP.}
  \label{fig:scenario1_20osf} 
\end{figure}

It is interesting to see the effect of increasing the oversampling factor for Lasso and DOMP. Figure~\ref{fig:scenario1_20osf} corresponds to Scenario~1, with the change being that the oversampling factor for these two algorithms is increased. We observe an improvement in the performance of both algorithms in terms of estimation accuracy, with that for Lasso being especially significant. Comparing these results to those in Figure~\ref{fig:snr_scen12}-a, we see that the MSE performance of DOMP is marginally improved by increasing $\gamma$ from $10$ to $100$. This performance plateau shows a fundamental \emph{algorithmic limitation} of DOMP, and highlights the critical role of cyclic Newton refinements in NOMP. In other words, the performance limitation of DOMP is not just due to the off-grid error, but also a consequence of making ``hard-decisions'' at each iteration. The computational complexity of DOMP is insensitive to oversampling factor. For example, computation time increases from $3.25$ to about $7.96$ seconds as we go from $\gamma = 20$ to $\gamma = 100$. 

In theory, Lasso is only limited by the oversampling factor $\gamma$, and as $\gamma \to \infty$, the result of Lasso converges to that of AST, as a 
consequence of the convergence of the corresponding atomic norms \cite{Recht1}. In practice, however, the performance of Lasso solved on a moderate size grid might be far from that of AST. Figure \ref{fig:scenario1_20osf} shows that, at large enough oversampling factors, Lasso approaches the CRB, but the computational cost becomes prohibitive. For example, the computation time of Lasso for $300$ runs, increases from $72.16$ seconds to $191.22$ seconds as we go from $\gamma = 20$ to $\gamma = 50$.


\subsection{Model order estimation}
\label{sec:model_order}
\begin{figure}[htbp]
\centering 
\subfigure{
\includegraphics[scale=0.18]{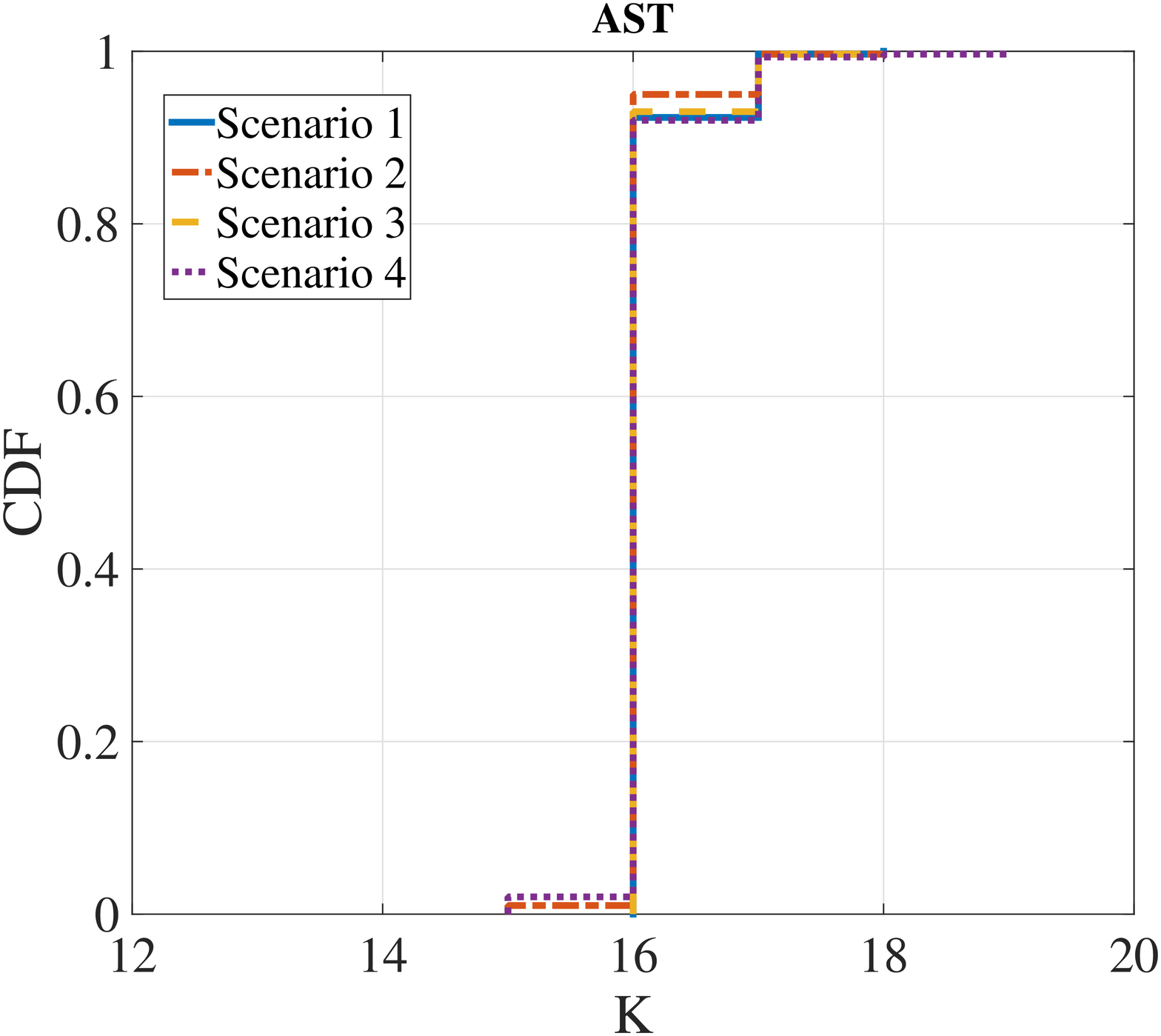}
} 
\hspace{.2in} 
\subfigure{
\includegraphics[scale=0.18]{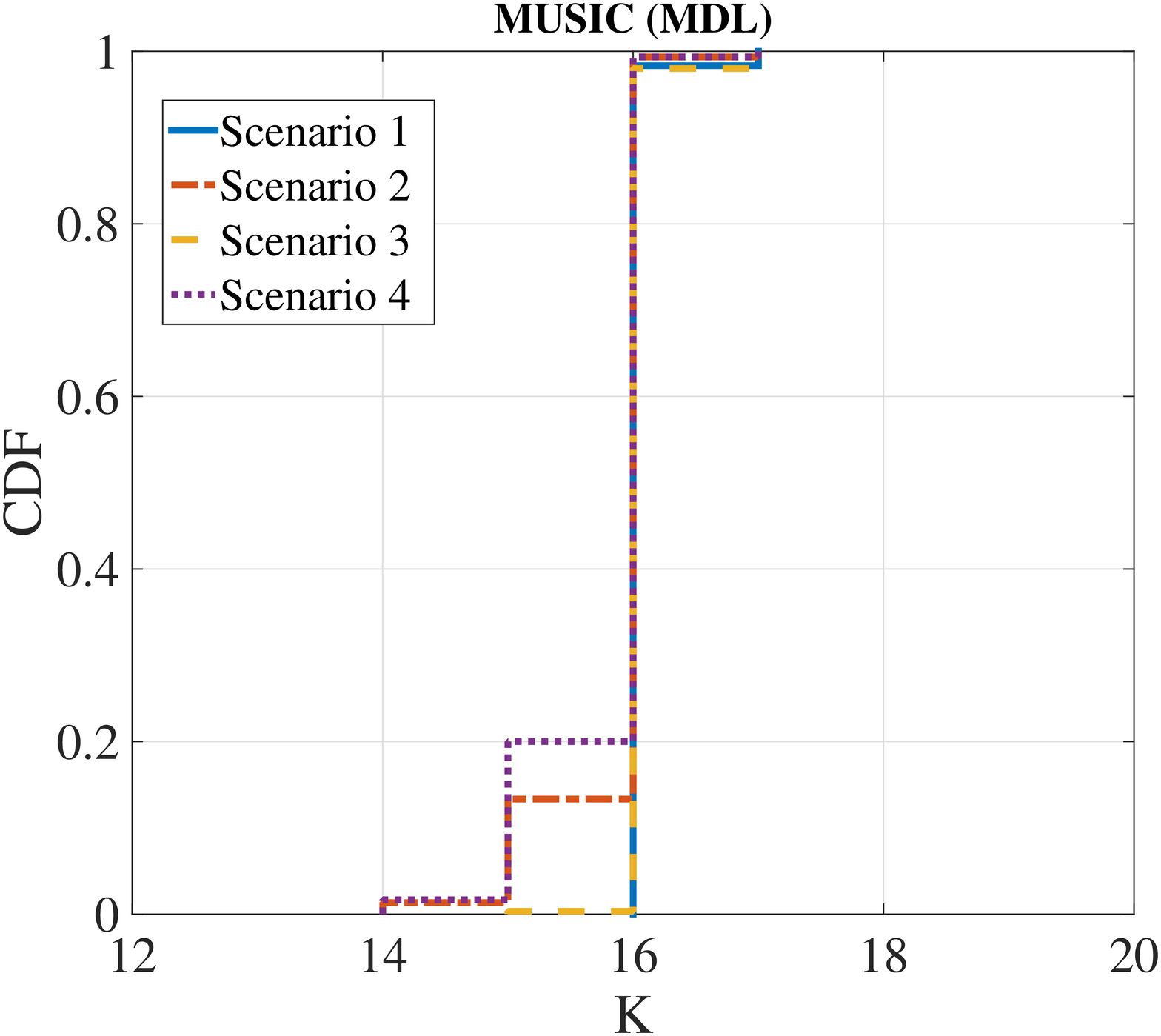}
}
  \caption{CDF of the estimates of the model order for AST and MUSIC.}
  \label{fig:SS_continuous} 
\end{figure}

\begin{figure}[htbp]
\centering 
\subfigure{
\includegraphics[scale=0.18]{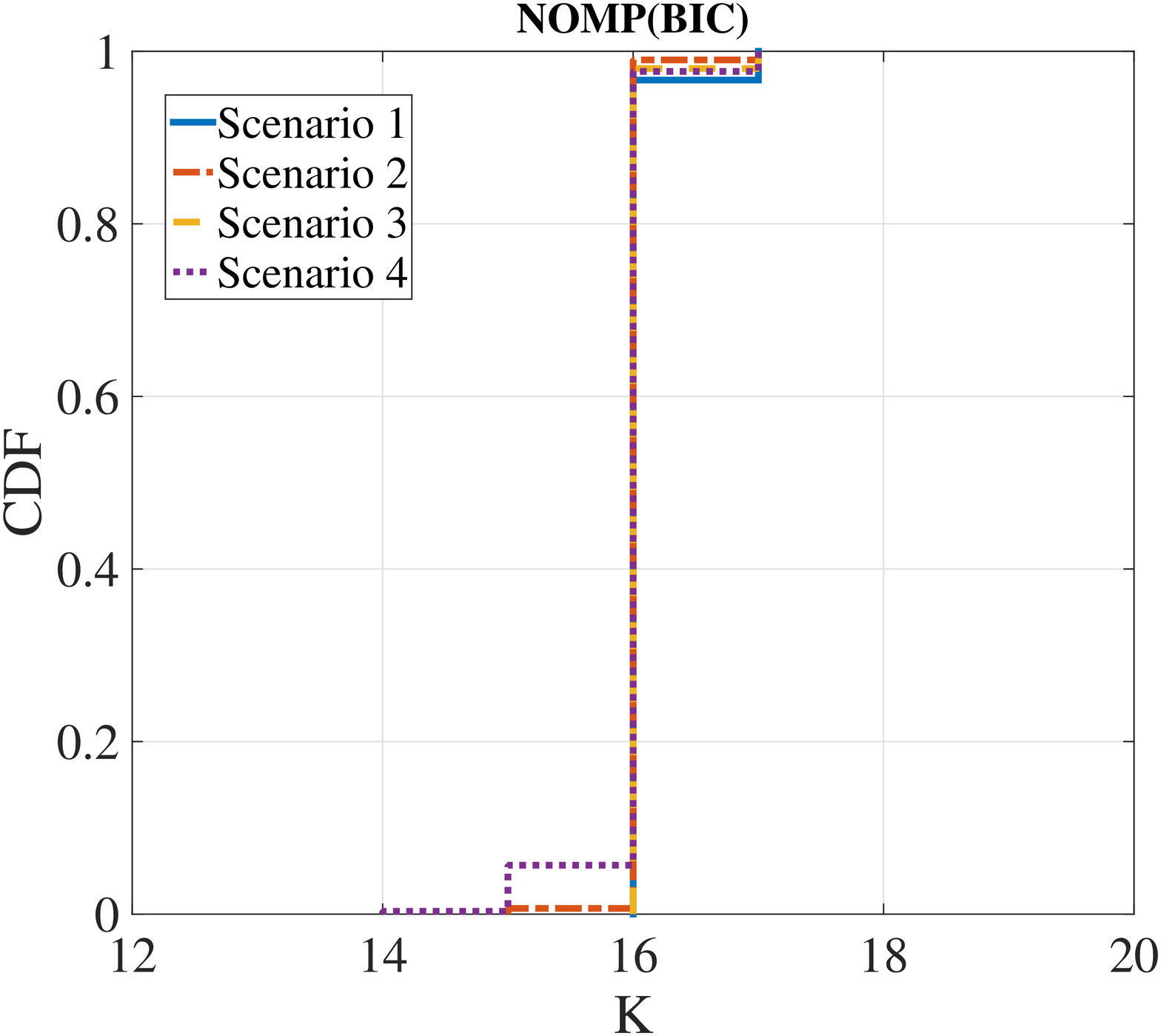}
} 
\hspace{.2in} 
\subfigure{
\includegraphics[scale=0.18]{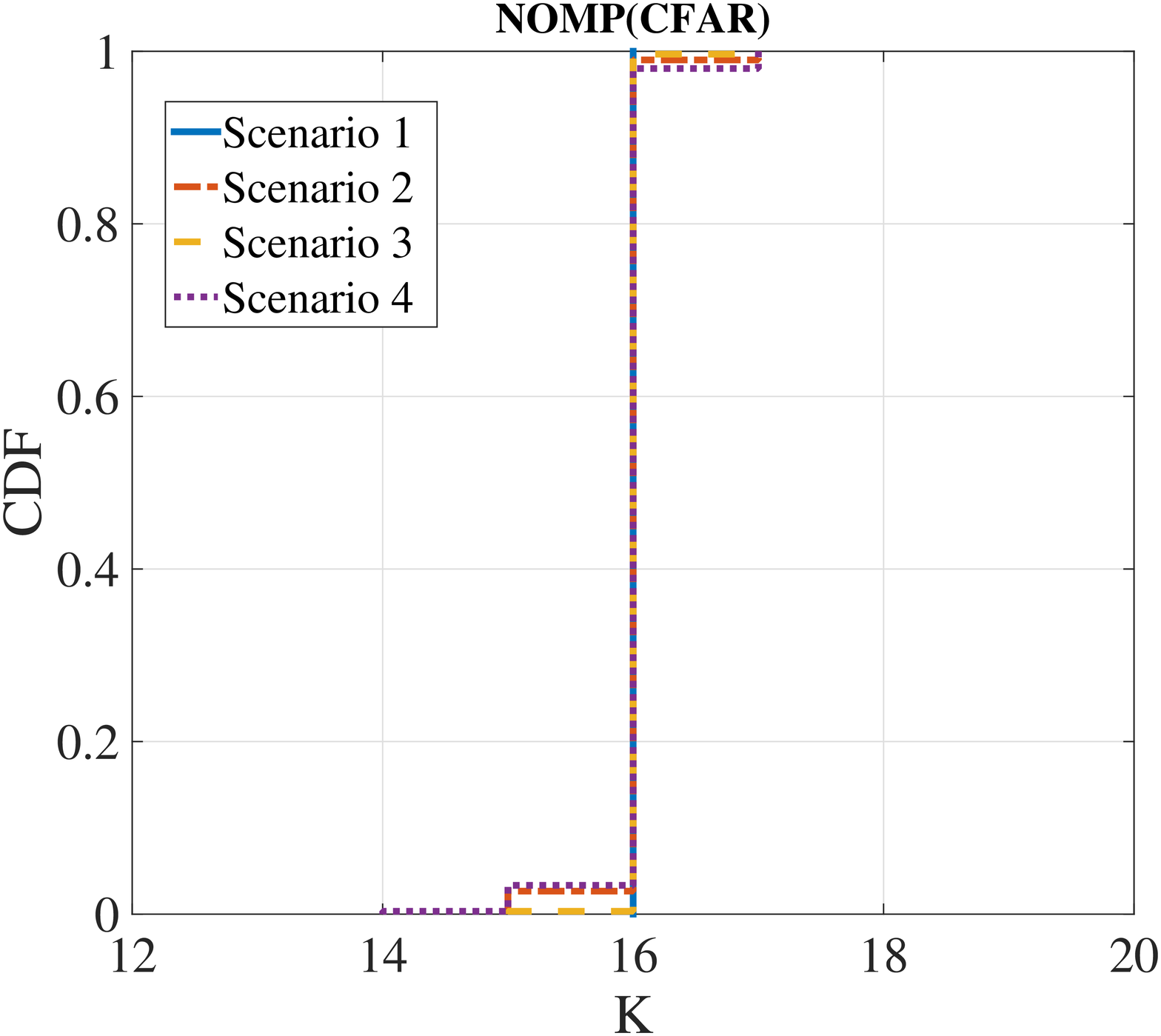}
}
  \caption{CDF of the estimates of the model order for NOMP using CFAR and BIC criteria.}
  \label{fig:SS_NOMP} 
\end{figure}

Estimating the model order $K$ (the true number of non-zero atoms in the mixture) has significant importance in sparse approximation. Here we examine the Cumulative Distribution Function (CDF) of the estimated model order by different algorithms. As shown in Figure \ref{fig:SS_continuous}, both AST and MUSIC perform  well, with MUSIC performance slightly degrading in scenarios with small $\minSep$.  Note that the mean of the distributions are very close to the truth (small bias), and they have a small spread around the mean (small variance). Figure \ref{fig:SS_NOMP} shows the model order estimates for NOMP using both CFAR and BIC-based stopping criteria. We see that both criteria are very accurate in all four scenarios.

\begin{figure}[htbp]
\centering 
\subfigure{
\includegraphics[scale=0.18]{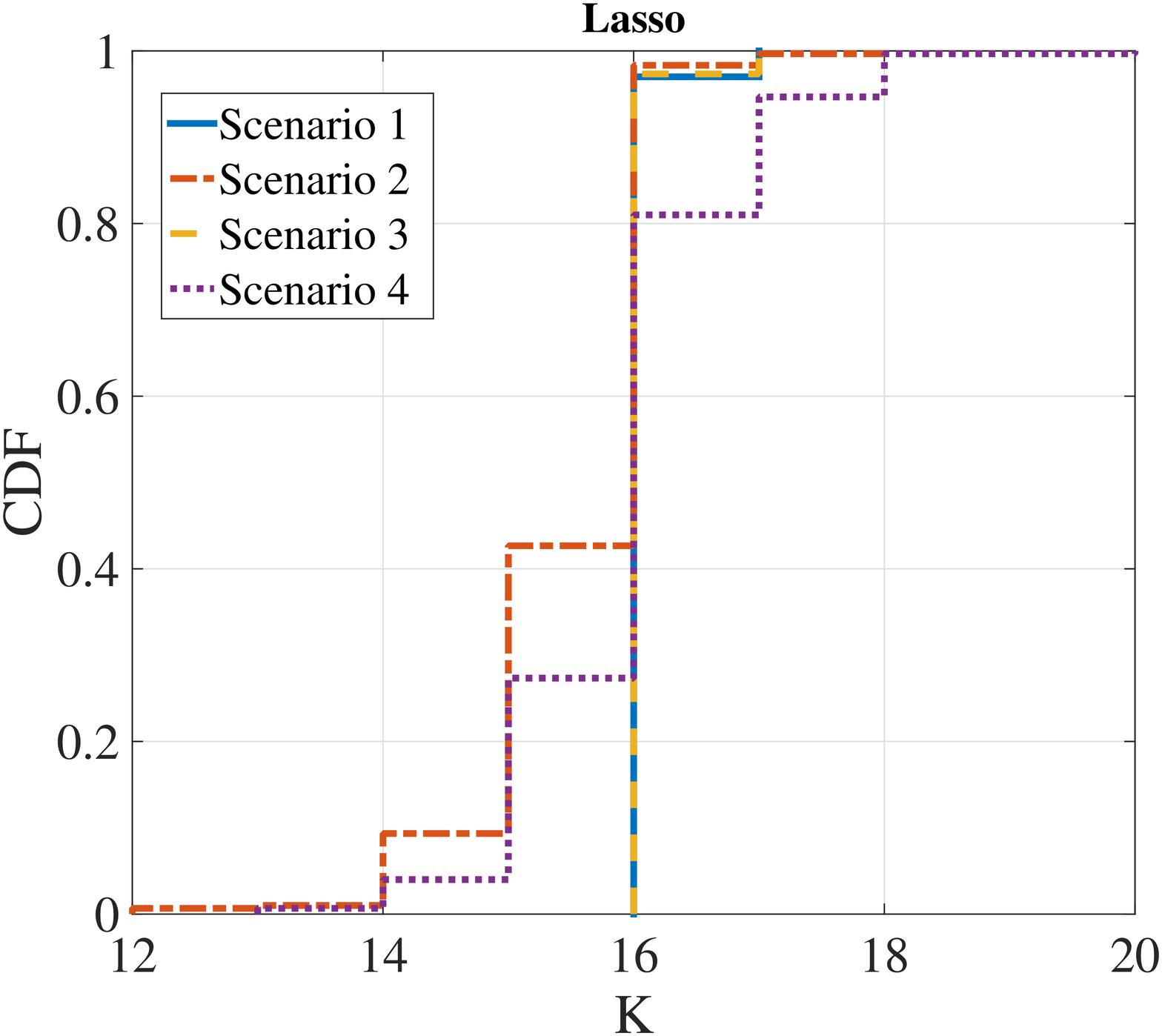}
} 
\hspace{.2in} 
\subfigure{
\includegraphics[scale=0.18]{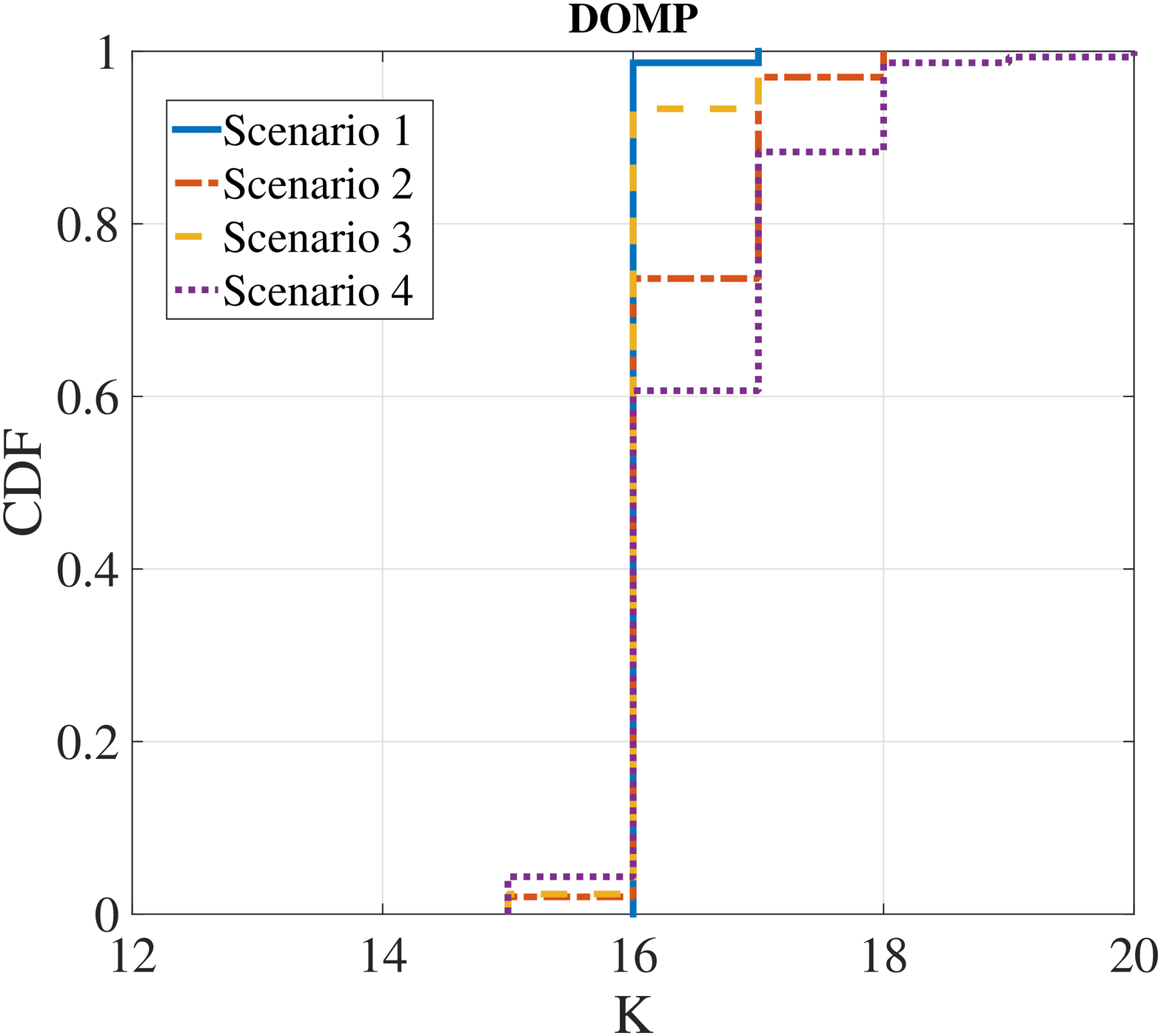}
}
  \caption{CDF of the estimates of the model order for Lasso and DOMP with $\gamma = 20$.}
  \label{fig:SS_discrete} 
\end{figure}

On the other hand, Lasso and DOMP perform poorly in estimating the model order, especially in scenarios with small $\minSep$. As shown in Figure \ref{fig:SS_discrete}, DOMP tends to overestimate the model order when some of the frequencies are placed too closely. This is again the result of a fundamental algorithmic limitation.  DOMP does not allow for correcting errors that have happened in the previous iterations; instead, it tries to explain the residual energy by overestimating the number of non-zero atoms.

In Scenario 2, Lasso tends to underestimate the model order. The main reason is that for two closely spaced frequencies, Lasso generates two overlapping clusters of estimated frequencies, which are later replaced by a single frequency by our \emph{clustering} algorithm. On the other hand, if we do not employ clustering, Lasso significantly overestimates the model order.

\section{Extensions of the Algorithm} \label{sec:extensions}
In this section we point out an immediate extension of NOMP algorithm. Specifically, we can replace the manifold of sinusoids $\{\bx(\omega)~:~\omega \in [0,2\pi)\}$ by \mbox{$\{A\bx(\omega)/\norm{A\bx(\omega)}~:~\omega \in [0,2\pi)\}$}, where $A\in \mathbb{C}^{M\times N}$ is a known measurement matrix. This is motivated by the following measurement setup:
\begin{equation}
\by = \sum_{l=1}^{K} g_l A \bx(\omega_l).
\end{equation}

\subsection*{Compressive measurements}
\begin{figure}[htbp]
\centering
\includegraphics[scale=0.19] {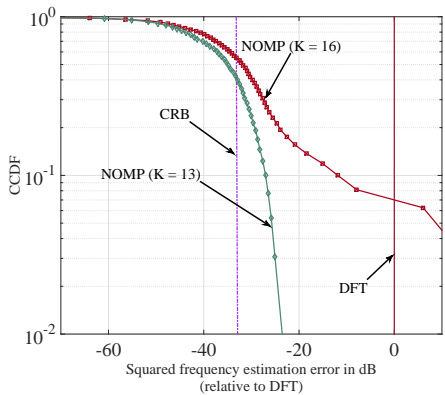}
  \caption{CCDF of the frequency MSE for Scenario 1 with Compressive measurements.}
  \label{fig:compressive} 
\end{figure}

We consider a compressive measurement model in which the number of measurements $M \ll N$. As in the bulk of literature on compressive sensing, we assume that the elements of $A$ are chosen i.i.d from appropriate zero-mean distributions (with variance conveniently scaled to $1/N$ such as $\text{Uniform}\{\pm 1/\sqrt{N}\}$, $\text{Uniform}\{\pm 1/\sqrt{N}, \pm j/\sqrt{N}\}$, $\mathcal{N}(0,1/N),$ etc.,) so that certain concentration results hold. It has been shown in \cite{Dinesh_TSP} that when $A$ satisfies certain isometry conditions (related to the estimation problem at hand), the CRB and ZZB are approximately preserved for compressive estimation, except for an SNR degradation of $M/N$ due to randomly projecting down the signal to a smaller space. The number of compressive measurements needed to estimate continuous frequencies scales as $M = \mathcal{O}(K\log N)$ \cite{Dinesh_TSP}.
In order to get concrete numerical intuition, we set $M = N/4 = 64$ with the elements of $A$ chosen uniformly and independently at random from $\{\pm 1/\sqrt{N}, \pm j/\sqrt{N}\}$ and run NOMP with atoms $\{ \mathbf{s}(\omega) = A\bx(\omega)/\norm{A\bx(\omega)}: \omega\in [0,2\pi) \}$. We consider Scenario~1 with $R_c = 3$, with number of sinusoids set to $K=13$ and $K=16$. Our algorithm approaches the CRB in the setting where $K=13$, whereas we incur large estimation errors when $K=16$; see Figure~\ref{fig:compressive}. The large estimation errors for $K=16$ occur because the compressive measurement matrix $A$ fails to preserve the structure of the estimation problem: $M=64$ compressive measurements is too few for $K=16$ sinusoids.  


\section{Conclusions} \label{sec:conclusions}

We have shown that NOMP is fast and near-optimal for frequency estimation in AWGN. It performs better than classical methods such as MUSIC, and more recent convex optimization based methods such as AST, in terms of both estimation accuracy and run time.
The algorithm uses a fundamental element of OMP, ensuring that the residue is orthogonal to the signal space spanned by the current
set of frequencies.  However, NOMP avoids the error floors of naively discretized OMP by refinement over the continuum.
Specifically, it searches for a signal subspace in the ``neighborhood'' of our current signal space which can better explain the observed measurements. 
The algorithm has a natural ``decision feedback'' interpretation in which it gives the already detected sinusoids a chance to adjust their frequencies in light of new evidence which is presented in the form of an updated residue after we add the next sinusoid.

We believe that the ideas underlying NOMP are broadly applicable to problems of sparse approximation over continuous
dictionaries, but further work is required to justify this assertion. An open problem is to go beyond the pessimistic convergence
rate estimates provided here by analytically quantifying the benefits of refinement.

\section{Acknowledgment}
This work was supported in part by Systems on Nanoscale Information fabriCs (SONIC), one of the six SRC STARnet Centers, sponsored by MARCO and DARPA.

\section*{Appendix I: Bayesian Information Criteria}
The BIC balances an increase in the likelihood with the number of parameters used to achieve that increase. Namely,
\begin{equation}
\text{BIC} = 2\ln{L_2(\theta_1,\dots,\theta_{m_2})\over L_1(\phi_1, \dots, \phi_{m_1})} - (m_2 - m_1) \ln (N),
\end{equation}
where $L_1(\phi_1, \dots, \phi_{m_1})$ and $L_2(\theta_1,\dots,\theta_{m_2})$ are likelihood functions and $\{\phi_{i}\}_{i=1}^{m_1}$ and $\{\theta_{j}\}_{j=1}^{m_2}$ are their corresponding parameters. For our measurement model, $\by = \sum_{l=1}^{K}g_{l}\bx(\omega_l) + \bz$, where $\bz\sim \mathcal{N}(\mathbf{0},\sigma^2\mathbb{I}_{N})$, we have 
$$
L(\{g_l,\omega_l\}_{l=1}^{K}) = {1\over(\pi\sigma^2)^{N}}\exp(-||\by_r||^2/\sigma^2),
$$
where $\by_r = \by - \sum_{l=1}^{K}g_{l}\bx(\omega_l)$ is the residual. Therefore, the BIC criterion will be as follows:
\begin{equation}
\text{BIC} = {2\over \sigma^2} \Delta||\by_r||^2 - 2\ln(N),
\end{equation}
where $\Delta ||\by_r||^2 = ||\by_r(old)||^2 - ||\by_r(new)||^2$ is the reduction in residual energy after detecting a new sinusoid. If \mbox{$\text{BIC}>10$}, it is a strong evidence that the most recent reduction in the residual energy corresponds to a newly detected sinusoid. Therefore, when $\text{BIC}<10$, we stop the algorithm.

\section*{Appendix II: \CRB}
In this Appendix we first review the \CRB~  \cite{CRLB} for an unbiased estimator in a general setting, then specialize to the frequency estimation problem. Let $\ba \in \mathbb{R}^{\lambda}$. If $\ba^T \hat{\theta}(\by)$ is an unbiased estimator of  $\ba^{T}\theta$, then the variance of the estimator given by $\mathbb{E}_{\by|\theta} \left[ \left(\ba^T \hat{\theta}(\by) - \ba^{T}\theta\right)^2 \right]$ is lower bounded by $\ba^T F^{-1}(\theta) \ba$, where $F(\theta)$ is the Fisher Information Matrix. The $(m,n)^{\text{th}}$ element of $F(\theta)$ is given by
\begin{equation} \label{eq:CRB_general}
F_{m,n}(\theta) = \mathbb{E}_{\by|\theta} \left\{ {\partial \ln p(\by|\theta)\over \partial \theta_m}  {\partial \ln p(\by|\theta)\over \partial \theta_n}  \right\}.
\end{equation}
For parameter estimation in additive white Gaussian noise i.e. $\by = \mathbf{s}(\theta) + \bz, ~ \bz \sim \mathcal{CN}(0,\sigma^2 \mathbb{I})$, equation (\ref{eq:CRB_general}) simplifies to,
\begin{equation} 
F_{m,n}(\theta) =  {2\over \sigma^2} \Re \left\{  \left({\partial \mathbf{s}(\theta)\over \partial \theta_m}\right)^H {\partial \mathbf{s}(\theta)\over \partial \theta_n}  \right\}.
\end{equation}
For our frequency estimation problem in (\ref{eq:meas_model}), $\theta$ is the vector of all  parameters, namely $\{|g_l|, \angle g_l, \omega_l : l = 1, \dots , K\}$, and $\mathbf{s}(\theta) = \sum_{l=1}^{K} |g_l| e^{j\angle g_l} \bx(\omega_l)$.  We form $F(\theta)$ for this measurement model then choose those diagonal elements of $F^{-1}(\theta)$ that correspond to the frequencies $\{\omega_l : l=1,\dots,K\}$. 

\bibliographystyle{IEEEtran}
\bibliography{Ref}

\end{document}